\documentclass[sigconf]{acmart}

\AtBeginDocument{%
  }

\acmPrice{15.00}

\acmSubmissionID{3592}

\copyrightyear{2024}
\acmYear{2024}
\setcopyright{rightsretained}
\acmConference[IUI '24]{29th International Conference on Intelligent User Interfaces}{March 18--21, 2024}{Greenville, SC, USA}
\acmBooktitle{29th International Conference on Intelligent User Interfaces (IUI '24), March 18--21, 2024, Greenville, SC, USA}
\acmDOI{10.1145/3640543.3645166}
\acmISBN{979-8-4007-0508-3/24/03}

\usepackage[section]{placeins}  

\usepackage{fontawesome}

\usepackage{multirow} 
\usepackage{tabularx} 
\usepackage{graphicx}
\usepackage{wrapfig}  

\usepackage{tikz} 
\usepackage{subfig}   

\newcommand*\darkcircled[1]{\tikz[baseline=(char.base)]{
            \node[shape=circle,draw=blue!10!gray,fill=blue!10!gray,inner sep=0.5pt, line width=0.5pt, text=white, font=\footnotesize] (char) {#1};}}

\newcommand{\EB}{\textcolor[RGB]{163, 128, 185}}
\newcommand{\Com}{\textcolor[RGB]{224, 161, 66}}
\newcommand{\Ho}{\textcolor[RGB]{81, 166, 171}}
\newcommand{\ExA}{\textcolor[RGB]{135, 204, 76}}

\newcommand{\higherthan}{\textcolor[RGB]{74, 147, 255}}
\newcommand{\lowerthan}{\textcolor[RGB]{250, 129, 129}}
\newcommand{\closeto}{\textcolor[RGB]{96, 98, 102}}

\begin{document}

\title{BiasEye: A Bias-Aware Real-time Interactive Material Screening System for Impartial Candidate Assessment}


 \author{Qianyu Liu}
 \email{liuqy@shanghaitech.edu.cn}
 \orcid{0009-0006-0212-1318}
 \affiliation{%
   \institution{School of Information Science and Technology, ShanghaiTech University}
   \city{Shanghai}
   \country{China}}

 \author{Haoran Jiang}
 \email{jianghr@shanghaitech.edu.cn}
 \orcid{0009-0009-5717-4208}
 \affiliation{%
   \institution{School of Information Science and Technology, ShanghaiTech University}
   \city{Shanghai}
   \country{China}}

 \author{Zihao Pan}
 \email{panzh@shanghaitech.edu.cn}
 \orcid{0009-0009-1074-906X}
 \affiliation{%
   \institution{School of Information Science and Technology, ShanghaiTech University}
   \city{Shanghai}
   \country{China}}

 \author{Qiushi Han}
 \email{hanqsh@mail2.sysu.edu.cn}
 \orcid{0009-0003-5455-579X}
 \affiliation{%
   \institution{School of Artificial Intelligence, Sun Yat-sen University}
   \city{Zhuhai}
   \country{China}}

 \author{Zhenhui Peng}
 \email{pengzhh29@mail.sysu.edu.cn}
 \orcid{0000-0002-5700-3136}
 \affiliation{%
   \institution{School of Artificial Intelligence, Sun Yat-sen University}
   \city{Zhuhai}
   \country{China}}

 \author{Quan Li}
 \authornote{The corresponding author.}
 \email{liquan@shanghaitech.edu.cn}
 \orcid{0000-0003-2249-0728}
 \affiliation{%
   \institution{School of Information Science and Technology, ShanghaiTech University, and Shanghai Engineering Research Center of Intelligent Vision and Imaging, China}
   \city{Shanghai}
   \country{China}
 }

\renewcommand{\shortauthors}{Liu et al.}

\begin{abstract}
  In the process of evaluating competencies for job or student recruitment through material screening, decision-makers can be influenced by inherent cognitive biases, such as the screening order or anchoring information, leading to inconsistent outcomes. To tackle this challenge, we conducted interviews with seven experts to understand their challenges and needs for support in the screening process. Building on their insights, we introduce \textit{BiasEye}, a bias-aware real-time interactive material screening visualization system. \textit{BiasEye} enhances awareness of cognitive biases by improving information accessibility and transparency. It also aids users in identifying and mitigating biases through a machine learning (ML) approach that models individual screening preferences. Findings from a mixed-design user study with 20 participants demonstrate that, compared to a baseline system lacking our bias-aware features, \textit{BiasEye} increases participants' bias awareness and boosts their confidence in making final decisions. At last, we discuss the potential of ML and visualization in mitigating biases during human decision-making tasks.
\end{abstract}

\ccsdesc[500]{Human-centered computing~Human computer interaction (HCI)}
\ccsdesc[300]{Human-centered computing~Visualization}
\ccsdesc[200]{Human-centered computing~User studies}

\keywords{bias-aware design, inconsistent decision, raise bias awareness, material screening in holistic review}


\maketitle

\section{Introduction}

\par The process of material screening during admissions plays a vital role in the intricate decision-making process for both college enrollment and corporate recruitment. Typically, this process involves independent reviews of various segments of the applicant's materials, resulting in a multidimensional assessment of their qualifications. Subsequently, reviewers record key points on a decision sheet for each application~\cite{Sukumar:2018:Pecan}.

\par Application materials encompass a diverse range of documents, including personal resumes, additional certifications and letters of recommendation, among others. Given the substantial volume of applications, various automated techniques have emerged to assist in systematically and efficiently extracting and storing information. These techniques include academic exploration~\cite{Gaur:2021:Semi,Swami:2022:Resume} and commercial solutions such as \textit{Daxtra}\footnote{http://www.daxtra.cn/} and \textit{Bello AI}\footnote{https://www.belloai.com/}. In terms of material screening, computer programs can provide a more objective and consistent assessment method based on predefined criteria~\cite{Lai:2016:CMA}. They are also employed to achieve diversity in candidate selection
~\cite{Gilbert:2021:Equitable}. However, these automated methods cannot comprehensively evaluate an applicant's personality and potential, nor can they fully grasp the complexities of background information. Human reviewers, on the other hand, excel at flexible adaptation~\cite{Kahneman:2011:Thinking}. Still, they are susceptible to \textbf{cognitive biases} stemming from perceptual illusions, false memories, logical fallacies and cognitive errors~\cite{Kahneman:2011:Thinking}. These biases are inherent in human perceptual and intuitive decision-making processes. While efforts can be made to identify and mitigate these biases, they cannot be entirely eliminated~\cite{Kahneman:2011:Thinking}. Furthermore, cognitive biases can be exacerbated by factors such as decision fatigue~\cite{Pignatiello:2020:Decision} and choice overload~\cite{Chernev:2015:Choice}.

\par During material screening, reviewers suffer from cognitive biases stemming from several challenges. These challenges also underscore the difficulty of raising awareness about and mitigating these biases in the decision-making process. First, \textbf{the lengthy and intermittent screening process can lead to \textit{recency bias}\footnote{The tendency to be excessively affected by the pattern of recent data.}}~\cite{Sinha:2022:Personalized}, as memory of earlier assessments fades over time, and decision-making criteria can be erratic (\textbf{$\mathcal I1$}). Second, \textbf{certain attributes of applicants can trigger ``\textit{halo}'' or ``\textit{horns}'' effect\footnote{The Halo/Horns effect is the idea that one's perception of someone is positively/negatively influenced by his/her opinion of that person's related traits.}}, hindering reviewers from providing unbiased assessments of other traits~\cite{Forgas:2016:Halo}. For instance, during the initial assessment of academic factors, reviewers might encounter an outstanding achievement, like a perfect math grade. This initial impression can lead to an \textit{anchoring bias}~\cite{Tversky:1974:Judgment}, where reviewers may expect equally exceptional performance in other areas then potentially lead to a biased evaluation of overall aptitude (\textbf{$\mathcal I2$}). Third, \textbf{balancing multiple admission goals including inclusivity and selectivity is challenging} due to memory limitations and cognitive workload. Reviewers may fall prey to the \textit{contrast bias}~\cite{Simonsohn:2013:Daily}, where their judgments are influenced by the scores given to the adjacent applicants (\textbf{$\mathcal I3$}). Lastly, forementioned challenges necessitate \textbf{inevitable revisions in the material screening process.} Reviewers often need to manually revise scores by reopening applicant pages, which can be challenging due to memory issues and \textit{confirmation bias}\footnote{The tendency to favor information that supports existing beliefs while disregarding contradictory evidence.}~\cite{Burke:2005:Improving} when revisiting applications consciously (\textbf{$\mathcal I4$}).

\begin{figure*}[h]
    \centering
    \vspace{-5mm}
    \includegraphics[width=\linewidth]{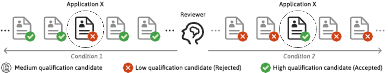}
    \vspace{-6mm}
    \caption{An illustration of the contrast bias emerges in sequential material screening tasks. In such scenarios, condition of adjacent application materials can influence a reviewer's assessment, resulting in reviewers making inconsistent judgments about the same application X under varying conditions.}
    \label{fig:contrastBias}
     \vspace{-3mm}
\end{figure*}

\par Artificial Intelligence (AI) approaches, while incapable of fully replacing human decision-making in college admissions, serve as valuable tools in addressing and mitigating various cognitive biases. Previous studies in this domain can be classified into several key categories based on the life cycle of bias~\cite{Croskerry:2013:Cognitive}: 1) \textit{Prevent}. Preventative training approaches~\cite{Capers:2017:Implicit,Fischhoff:1975:TSCo} aim to explicitly raise awareness of bias, although they can impose a significant cognitive burden on users. Procedural interventions, on the other hand, integrate bias awareness into the decision-making workflow by enhancing information transparency~\cite{Zhu:2022:Bias} or providing relevant information to reviewers. 2) \textit{Discover} and 3) \textit{Locate}. Researchers have developed models to detect biases in real-time~\cite{Wall:2017:Warning,Melton:2022:Using,Echterhoff:2022:AI-Moderated} and communicate these biases to users through visual elements~\cite{Narechania:2021:Lumos,Wall:2022:Left}, based on the definitions of different cognitive biases. 4) \textit{Mitigate}. Mitigation strategies and algorithms can be introduced based on machine learning methods~\cite{Echterhoff:2022:AI-Moderated,Akl:2016:DIMCBHD,Sinha:2022:Personalized} or visual approaches~\cite{Wall:2019:TDSMCBV,Chuanromanee:2022:ACSVSMCB,sukumar:2017:holistic,sukumar:2018:visualization}, offering promising avenues to reduce cognitive bias. 
However, existing modeling methods often \textbf{target specific defined biases, neglecting the interaction between cognitive biases}~\cite{Kahneman:2011:Thinking} (research gap \textbf{RG1}). For example, anchoring bias from the earliest applications and recency bias from the recent applications may affect next decisions in the same or opposite way. Regardless of the type of bias, they can lead to inconsistent decision outcomes, as illustrated in \autoref{fig:contrastBias}. In college admissions, these inconsistent screening results may conflict with the principle of \textit{individual fairness}~\cite{Echterhoff:2022:AI-Moderated}, where individuals may apply different criteria at different stages of a decision task, resulting in instances with similar characteristics being treated disparately. Previous studies on fairness and diversity in college admissions~\cite{Lai:2016:CMA,Gilbert:2021:Equitable} primarily focus on the rationality of final admission outcomes, \textbf{overlooking personal inconsistent outcomes and the individual material screening process} (\textbf{RG2}). 
Regarding visualization approaches, previous research~\cite{Sukumar:2018:Pecan,sukumar:2017:holistic,sukumar:2018:visualization} has demonstrated the potential of visual and interactive strategies to enhance human decision-making theoretically. Nonetheless, the \textbf{integration of AI methods and visualization strategies to address cognitive biases was infrequent}, and there was limited assessment of their combined effectiveness in practical applications (\textbf{RG3}).

\par To explore the factors contributing to inconsistent decision-making outcomes and the needs of reviewer for a feasible screening system, we conducted interviews with seven experienced reviewers from various academic disciplines in local universities. Based on six findings obtained from these interviews (\autoref{sec:screenFact}), we identified four primary challenges regarding details about four cognitive biases (introduced as \textbf{C1-C4} in \autoref{sec:Challenge}). In light of our literature review and identified challenges, in \autoref{sec:Mstep} we devised a four-step pipeline, \textit{PREVENTING} $\rightarrow$ \textit{DISCOVERING} $\rightarrow$ \textit{LOCATING} $\rightarrow$ \textit{MITIGATING} (\textbf{RG1}), applying to inconsistent decision-making that results from any cognitive biases, along with five essential design requirements for developing an effective system. Subsequently, we conceptualized and developed \textit{BiasEye}, a bias-aware real-time interactive material screening visualization system. \textit{BiasEye} served the purpose of prompting, tracking, and scrutinizing individual decision-making (\textbf{RG2}) during screening process in accordance with the four-step pipeline. The system’s backend employed \textit{ChatGPT-4} to extract features from application materials and models individual screening preferences through a machine learning (ML) approach (\textbf{RG3}). On the frontend, \textit{BiasEye} offered a side view that visualizes statistics for a group of applications, with each application being highlighted, as well as a summary page for retrospective decision inspection and adjustment. To assess the utility and effectiveness of \textit{BiasEye}, we conducted a mixed-subjects user study involving $20$ participants (\textbf{RG3}). The study provided strong support for the enhanced usefulness and effectiveness of \textit{BiasEye} compared to a baseline system and any combination of baseline systems with the addition of the side view or a summary page. Notably, \textit{BiasEye} helped participants implicitly reduce their inconsistent screening results without introducing or suggesting cognitive bias explicitly. Although the additional design elements increased cognitive load, participants reported increased confidence in their screening results' perceived reasonability and consistency. Furthermore, the system aided participants in better establishing their evaluative criteria, resulting in more concentrated scoring for high-quality applicants within the same group. Additionally, our observations indicated that \textit{BiasEye} facilitated participants' understanding and explanation of the model predictions. The presence of convincing evidence played a crucial role in their final level of trust in the system’s predictions. Building upon our findings, we put forth several design implications for future developments in material screening systems. The main contributions of this study include:
\begin{itemize}
\item In a formative study with seven participants, we identified challenges leading to cognitive biases in material screening and proposed a four-step pipeline to address them.
\item We developed \textit{BiasEye}, an interactive screening system that models decision preferences and mitigates bias.
\item A user study with 20 participants evaluated \textit{BiasEye}'s usability, effectiveness, and impact on behavior, workload, and confidence in screening outcomes.
\end{itemize}

\section{Related Work}

\subsection{Material Screening During the College Admission Review Process}
\par In college admissions, the holistic review approach has been widely recognized and explored across various domains~\cite{lucido:2014:admission,thresher:1966:college}. A critical component of this process is material screening, which occurs after the application submission and precedes the committee meeting. Talkad Sukumar et al.~\cite{Sukumar:2018:Pecan} conducted an in-depth study on the holistic review process employed by American universities, with a specific focus on aspects related to human-computer interaction and technical support. As Sukumar et al. described, application reviewers are entrusted with a pivotal phase known as \textit{Material Screening}, where reviewers draw upon their expertise and apply a predefined set of criteria, aligned with the university's mission and objectives, to evaluate applications. This comprehensive evaluation encompasses a wide array of factors gleaned from the materials submitted by applicants, including a student's high school background, family history, encountered challenges, as well as both academic and non-academic achievements, such as community service and special talents~\cite{Sukumar:2018:Pecan}. The material screening process is inherently subjective and intricate. It requires reviewers to assess applicants within the broader context of their individual backgrounds and life experiences. Rather than following a rigid, predefined protocol, reviewers rely on flexible personal heuristics, however such subjectivity would inadvertently introduce systematic errors or biases~\cite{Tversky:1974:Judgment} such as anchoring and confirmation bias~\cite{Sukumar:2018:Pecan}. This study (\autoref{sec:Challenge}) will explore how four task challenges associated with four prominent biases affect the screening process and examine the tools and methodologies employed by reviewers, providing valuable insights into this crucial stage of college admissions.

\subsection{Human Bias Detection and Mitigation}
\par The college admissions screening process has low validity, limiting the ability to discern patterns and develop accurate intuitions, making experts prone to cognitive biases like anchoring bias~\cite{Cho:2017:TAEDMVA,Selim:2021:ABCDMENI,Valdez:2018:PAEV,Rudiger:2016:Cognitive}, attraction effect~\cite{Dimara:2019:MAEV}, and confirmation bias~\cite{Burke:2005:Improving}. These biases have been extensively studied and categorized in comprehensive taxonomies~\cite{Wall:2017:Warning,Dimara:2020:ATBTCBIV,Nussbaumer:2016:AFCBDFVAE,Pohl:2014:SCBMVA}. Additionally, research shows that the order of presenting the same information can significantly influence decision-making~\cite{Akl:2016:DIMCBHD}, recent personal decisions can serve as anchors, leading to errors or inconsistencies when reviewing the same case~\cite{Echterhoff:2022:AI-Moderated}. Aligned with~\cite{Echterhoff:2022:AI-Moderated}, we advocate for \textit{individual fairness}, ensuring similar individuals are treated equitably while extends beyond addressing anchoring bias. In this study, we use ``\textbf{inconsistent}'' to describe situations where individual fairness is violated within the material screening process.

\par Detecting and mitigating cognitive bias is crucial in decision-making processes, and previous work falls into four distinct categories: \textbf{1) Preventing}. Some studies~\cite{Capers:2017:Implicit,Fischhoff:1975:TSCo} have focused on prevention by utilizing training approaches to raise awareness and discourage biased heuristics. However, relying solely on prior knowledge may not effectively mitigate biases and can impose cognitive burdens on users~\cite{Capers:2017:Implicit,Fischhoff:1975:TSCo}. Procedural interventions integrate bias avoidance into workflows without explicitly highlighting biases, such as increasing information transparency~\cite{Zhu:2022:Bias} and providing more relevant information to assess applicants, thereby improving the retrievability of relevant instances~\cite{Sukumar:2018:Pecan}. \textbf{2) Discovering}. Researchers have used machine learning and visual environments~\cite{Nussbaumer:2016:AFCBDFVAE,Sinha:2022:Personalized} to detect human biases, some have defined and measured bias indicators~\cite{Wall:2017:Warning,WallEmily:2019:AFSo}. This category is closely associated with the next: \textbf{3) Locating}. Studies such as~\cite{Wall:2022:Left} and~\cite{Narechania:2021:Lumos} visualized bias indicators within situational or peripheral view to pinpoint the source of bias. Echterhoff et al.~\cite{Echterhoff:2022:AI-Moderated} captured a reviewer's anchoring state using a probabilistic model to retrospectively locate biased decisions. \textbf{4) Mitigating}. Akl et al.~\cite{Akl:2016:DIMCBHD} developed strategies to reduce order-effects and enhance decision-making based on probability models. Visual methods, such as design spaces~\cite{Wall:2019:TDSMCBV} and simple visual representations~\cite{Chuanromanee:2022:ACSVSMCB}, have been proposed to mitigate cognitive bias. Researches~\cite{sukumar:2017:holistic,sukumar:2018:visualization} have demonstrated that implementing visualizations in the review process can automatically address cognitive biases, alleviating user concerns.

\par In this study, drawing inspiration from prior research, we have integrated a four-step pipeline into our material screening system. First, we present supplementary information and statistics related to applications to \textbf{prevent} cognitive bias. Next, employing machine learning techniques, we create dynamic models of real-time individual decision preferences based on a user's historical choices. Through our visualization design, users can \textbf{discover} and \textbf{locate} any inconsistencies in their decisions, ultimately helping them \textbf{mitigate} these inconsistencies conveniently.

\subsection{AI-Enhanced Approaches for Material Screening and Holistic Review Support}
\par Material screening serves as the crucial initial step in assessing a candidate's qualifications. To enhance efficiency and fairness, various methods have been developed to optimize procedures such the holistic admission process~\cite{JoubertChristel:2022:Spfa} and information ordering~\cite{Akl:2016:OIOSDCB}, or automate particular tasks such as resume screening~\cite{Ransing:2021:SRRSM}, assessment~\cite{Lai:2016:CMA}, and information extraction~\cite{Li:2021:AMRIEUBBC,Jia:2018:CNERBCBC}. Natural Language Processing (NLP)~\cite{Harsha:2022:ARSNLP} also has been used to detect and correct resume errors~\cite{Nasr:2019:AGSRUSTSM} and conduct rating classification while reducing human bias~\cite{Alamelu:2021:RVFNLP}.

\par Although automated screening can mitigate human bias, concerns about potential discrimination stemming from biased data or algorithms, including racial discrimination, have been raised~\cite{Noble:2021:The,Derous:2018:BDRS,Deshpande:2020:MDBABRF,Derous:2018:WYRTDMEBRS}. Initiatives like FairCVtest~\cite{Peña:2020:BMATFAR}, MANI-Rank~\cite{Cachel:2022:MRMAIGFCR}, and Gilbert et al.'s human-centered AI tool~\cite{Gilbert:2021:Equitable} aim to address these issues. However, it's important to note that while these methods automate parts of application material processing, they may not fully capture human review patterns or contextual nuances, limiting their use in holistic admissions reviews. Our approach uses machine learning as a supportive tool for human decision-making, adapting to individual reviewer preferences to personalize bias mitigation while retaining the final decision in the hands of the human reviewer.

\par Several software platforms, such as \textit{Slate}, \textit{Kira Talent}, and \textit{Submittable}~\cite{Submittable:2021:holistic}, support holistic review processes. The American Association of Medical Colleges (AAMC) also offers tools and principles for holistic review~\cite{Association:2021:Holistic}. Additionally, the College Board and Education Counsel jointly published a guide~\cite{coleman:2018:understanding} that includes a diversity metrics dashboard. Metoyer et al.~\cite{Metoyer:2020:Storytelling} explored group decision-making and integrated visual storytelling support into collaborative review for transparency and rigor. While these efforts focus on addressing cognitive bias and human decision-making in holistic admissions, our study centers on the material screening process prior to committee meetings. We aim to enhance bias awareness and promote self-reflection through machine learning and visualization in digital applications, building models of reviewers' personal decision preferences.

\section{Formative Study}
\par This study aims to help reviewers deal with inconsistent review decisions caused by cognitive biases. To achieve this, we conduct a formative study to understand reviewers' current practices and needs. These insights will inform the design requirements for a system tailored to this context.

\subsection{Participants and Procedure}
\par To comprehensively understand the current state-of-the-art material screening process, challenges faced by reviewers, and their expectations for screening systems, we conducted semi-structured interviews with seven experienced reviewers. The participants, with a mean age of 32.6 years (standard deviation 13.1), included four males and three females, offering diverse perspectives. Our objectives were twofold: to explore current practices and challenges in material screening and identify strategies to mitigate cognitive biases while enhancing efficiency and satisfaction. Interviewees represented various roles, including admissions officers, material reviewers, and interviewers, spanning academic and professional backgrounds in fields like Computer Science, Industrial Design, Entrepreneurial Finance, and FinTech. We developed the interview script through informal discussions with admission officers and reviewers. As outlined in \autoref{tab:interview}, participants discussed their screening procedures, experiences, and shared views on four cognitive biases: \textit{anchoring bias}, \textit{recency bias}, \textit{contrast bias}, and \textit{confirmation bias}, along with coping strategies and specific requirements within each scenario.

\par We used Braun and Clarke's six-phase thematic analysis framework~\cite{Guest2011:ThematicAnalysis} to analyze interview data. The analysis involved two researchers proficient in qualitative research methods. One researcher performed the initial coding of the data, while the other meticulously reviewed the codes to ensure accuracy and completeness.  Through iterative discussions, two authors reached a consensus on the summarizing statements at first, resolving potential ambiguities or conflicts. Next, they collaboratively identified six screening findings together, subsequently giving rise to four key challenge themes discussed in \autoref{sec:screenFact} and \autoref{sec:Challenge}, respectively. These insights informed the derivation of five design requirements, forming the foundation for a four-step strategy explained in \autoref{sec:Mstep}.

\begin{table*}[h]
\begin{tabular}{ll}
\hline
Category    & Question                                         \\ 
            \hline
Demographic & How many times have you participated in material screening or interviews for college admissions?                                 \\ 
            \hline
            & What is the overall flow of college admissions?                                                                                      \\
            & What policies/criteria has the admissions committee formulated?                                                                      \\
Procedures  & Who is qualified to be a reviewer? (How the Admissions Committee recruited the reviewers?)                                           \\
            & How the applications were distributed to reviewers?                                                                                  \\
            & How to evaluate an application comprehensively and decide admission results?                                                         \\
            & What are the functions of current screening system?    \\ 
            \hline
Prompt      & Have you ever overestimated or underestimated applications?                                 \\
            & How did you handle this problem and avoid the similar situation? \\ \hline
            & (1) Do you think that certain aspect of the applications will affect your assessment of their other\\
            & \quad\ aspects? \textbf{\texttt{[anchoring bias]}} \\
Scenarios   & (2) Do you think the review process is affected by time and memory? \textbf{\texttt{[recency bias]}}                                      \\
            & (3) Do you think the sequence order of applications may affect your assessment? \textbf{\texttt{[contrast bias]}}                     \\
            & (4) Did you objectively assess the shortcomings of applications when you have had a favorable     \\
            & \quad\ impression of them? \textbf{\texttt{[confirmation bias]}}\\ \hline
Expectation & What functions do you want to add or improve to the current screening/interview system?                                              \\ \hline
\end{tabular}
\caption{Interview with expertise reviewers.}
\label{tab:interview}
\vspace{-6mm}
\end{table*}

\subsection{Findings about the Current Material Screening Process} \label{sec:screenFact}
\par This section presents six key findings from our interviews about the current material screening process, comparing them with findings in~\cite{Sukumar:2018:Pecan}.

\par \textbf{\textit{Finding 1: Multiple rounds of material screening.}} Material screening has become more complex and time-consuming, with universities adopting a multi-round approach (E1, E2, E5), differing from the simplified approach in~\cite{Sukumar:2018:Pecan} where one reviewer was assigned per applicant. Moreover, reviewers encompass a spectrum of experience levels, ranging from senior assistant students acting as junior reviewers to professors serving as expert reviewers in each evaluation cycle. This approach achieves a dual objective of maintaining selectivity and inclusivity simultaneously. As E1 noted, ``\textit{A significant number of applications exist, and junior reviewers should screen out the underperforming ones, thus allowing expert reviewers to focus on the more competitive submissions.}'' Applicants undergo multiple reviews leading to interviews and committee meetings to finalize admissions.

\par \textbf{\textit{Finding 2: Multiple reviewers in each round.}} To mitigate the impact of personal preferences, each applicant is assigned to reviewers from various departments, and their scores are averaged to determine the effective score (E2, E4, E6, E7). According to E6, ``\textit{Reviewers possess their own preferences, and enabling reviewers with diverse backgrounds to assess the same applicants aligns with the objective of achieving a more diversified admissions process.}'' Similar to~\cite{Sukumar:2018:Pecan}, reviewers primarily handle applications from their respective or familiar regions but not exclusively so.

\par \textbf{\textit{Finding 3: Diverse admission expectations.}} As noted in~\cite{Sukumar:2018:Pecan}, reviewers are tasked with balancing diverse and inclusive admission goals with the school’s mission. Moreover, fair assessment is ensured by considering the average scores from at least three reviewers per round. Assuming a normal quality distribution, admission committee manages a large volume of applications by randomly distributing and sequencing them. As noted by E3, E4, and E7, reviewers are instructed to target a suggested mean score, mitigating aggregation errors.

\par \textbf{\textit{Finding 4: Flexibility in reviewer work schedules.}} Reviewers are provided with one to two weeks to autonomously accomplish their screening assignments, usually during breaks in their regular work and study schedules, as outlined in~\cite{Sukumar:2018:Pecan}. Reviewers have the flexibility to either assess a few applications daily during their spare moments or allocate a dedicated continuous time block to evaluate all applications (E1-7).
    
\par \textbf{\textit{Finding 5: Aggregating multi-dimensional assessments.}} Candidate assessment involves considering various dimensions such as educational background, academic and non-academic activities, and letters of recommendation~\cite{Sukumar:2018:Pecan}. Universities assign weights to each dimension for an overall score, rather than a single cumulative score. Furthermore, the admission office provides a list of competitions or awards for seamless integration into the scoring system, as E4 mentioned, ``\textit{This process has been automated recently as part of the system iteration.}''

\par \textbf{\textit{Finding 6: Outdated material screening system.}} As discussed in~\cite{Sukumar:2018:Pecan}, existing material screening systems are predominantly representational and lack interactivity. Reviewers navigate a list of applications, each with bundled PDF materials and a digital decision sheet for scoring and commenting. The application list shows screening progress, scores, and a submission button. While these systems provide basic functionalities like electronic storage and accessibility, they lack advanced features.

\subsection{Challenges in Material Screening Process} \label{sec:Challenge}
\par In this section, we will explore each challenge themes (\textbf{C1-C4}) by examining the fundamental characteristics (\textit{Finding 1$\sim$6}) of the material screening process, helping us identify potential cognitive biases in this phase.

\par \textbf{C1: Balancing workload and fairness in college admissions screening.} Despite the need for a holistic approach in college admissions (\textit{Finding 3}), the high volume of applications often restricts the time and energy reviewers can dedicate to each student, hindering thorough exploration and deliberation (E1, E3, E4, E6). The automated awards-to-score approach in \textit{Finding 5} may reduce some workload, but not all awards are listed, and subjective judgment based on experiential knowledge remains necessary. As E7 stated, aligning average students with score distribution requirements in \textit{Finding 3} is challenging, ``\textit{How do you come up with the boundary for those average students? It's a bit tricky, and honestly, I didn't know it right from the beginning.}'' The \textit{contrast bias}\cite{Simonsohn:2013:Daily} is particularly evident with intermediate qualifications, an outstanding applicant can overshadow others, and a series of subpar materials may lead to higher scores for an average applicant~\cite{Echterhoff:2022:AI-Moderated}.

\par \textbf{C2: The screening procedure can be quite time-consuming and frequently intermittent.} As highlighted in \textit{Finding 4}, the screening task is susceptible to interruptions and places a substantial memory burden on reviewers due to their constrained time (E4, E6) and the fragmented nature of their personal schedules (E1, E2, E3). Moreover, the influence of \textit{recency bias}~\cite{Sinha:2022:Personalized} prompts reviewers to base their decisions on applicants they've recently assessed (E5), resulting in a fluctuating personal evaluation criterion.
    
\par \textbf{C3: Reviewers might be susceptible to the allure of the halo effect\footnote{The Halo effect means one's perception of someone is positively influenced by his/her opinion of that person's related traits.}.} As \textit{Finding 5} and~\cite{Sukumar:2018:Pecan} suggest, an applicant's academic performance can anchor assessments of other dimensions (E1, E3, E4), validating the \textit{anchoring bias}~\cite{Tversky:1974:Judgment}. Furthermore this anchoring effect can positively or negtively manifest in various aspects. For instance, E3 remarked, ``\textit{This student possesses extensive experience and stands out among applicants. Excellent! I'm inclined to award extra points in every dimension.}'' Conversely, E4 expressed doubts, stating, ``\textit{Did his parents ghostwrite this self-introduction? Some phrases appear to be readily available online, suggesting a lack of sincerity, which raises concerns about the other achievements.}'' Anchoring bias can subtly influence reviewers using a heuristic approach to decision-making, resulting in unintentional inconsistent outcomes.

\par \textbf{C4: Reviewers struggle with inconvenient systems and lack guidance when making score adjustments.} Not discussed in~\cite{Sukumar:2018:Pecan}, initial lower scores~\cite{Bian:2022:Good} resulted from reviewers’ caution due to incomplete understanding. As E7 expressed, ``\textit{I acknowledge that this may seem unfair to students in the front. I'll proactively make adjustments, although I can't guarantee them.}'' As screening progressed, decision fatigue led to declining decision quality and preference for expedient or mean-score heuristics (E4). Inconsistent outcomes (C1-C3) necessitated revisions, with reviewers revisiting decisions repeatedly and verifying before final submissions. E1 emphasized iterative score revisions to ensure fairness, stating, ``\textit{It's essential to make adjustments, especially when more competitive performances are niticed further back. I need to lower those high scores at the front.}'' However, this was exhausting due to unreliable memory and the outdated system (\textit{Finding 6}). E6 suggested making retrospective assessment more intuitive and optimizing interaction beyond the current ``click to display'' method.

\vspace{-3mm}
\subsection{A Four-Step Pipeline and Design Goals} \label{sec:Mstep}
\par Drawing from relevant research and interview insights, we present a four-step pipeline to address four challenges and providing a foundation for system design \textbf{$\mathcal D1$}$\sim$\textbf{$\mathcal D5$}.

\par \textbf{Step 1: Preventing.} Humans may not consistently excel at repetitive tasks~\cite{Dix:2003:Human}, so enhancing screening quality, especially addressing \textbf{C1}, involves reducing the reviewer’s workload. \textbf{This means focusing on automating routine tasks, allowing reviewers to devote their attention to more complex subjective assessments ($\mathcal D1$).} To tackle \textbf{C1}, \textbf{$\mathcal D1$} includes preprocessing and gathering necessary information for screening applications, thus enhancing the retrieval of instances related to the availability heuristic~\cite{Sukumar:2018:Pecan}. Furthermore, automating repetitive judgments and actions through intuitive representation and simplified interaction is a practical strategy for \textbf{C4}.

\par \textbf{Step 2: Discovering.} Screening procedure and human cognitive process constraints can lead to biased decisions (\textbf{C1-C3})~\cite{Kahneman:2011:Thinking}. However, participants were either unaware of or underestimated bias impact in their assessments (E4, E7). With subjective criteria involving intangible, shifting factors, an objective approach is needed to help recognize and rectify irrational behavior. Our system aims to \textbf{facilitate reviewers' understanding and management of the screening process} (\textbf{$\mathcal D2$}), as well as \textbf{explicitly reveal screening preferences to uncover potential inconsistencies} (\textbf{$\mathcal D3$}).

\par \textbf{Step 3: Locating.} While the initial \textit{discovery} step offers an overview of the screening process, in-depth analysis is crucial to implement targeted strategies addressing inconsistencies from \textbf{C1-C3}. \textbf{$\mathcal D4$} involves \textbf{examining bias tendencies and evaluating specific bias instances}. Our system should provide transparent, comprehensive information for multifaceted material comparisons. Through interactive visualization, reviewers can identify inconsistent outcomes and make informed judgments, promoting fairness and objectivity in screening.

\par \textbf{Step 4: Mitigating.} The final key step in bias mitigation is score modification. The current system, mainly representational, lacks interactivity, burdening reviewers physically and cognitively when adjusting decisions (\textbf{C4}). Additionally, comparing numerous similar candidates for reasoned scores is challenging (E7). To address \textbf{C4}, our system should \textbf{enable quick adjustments and provide reasonable score recommendations} (\textbf{$\mathcal D5$}) to ease reviewers' bias concerns.  Meanwhile, interactive visualization is a promising way to enhance assessment efficiency and effectiveness.

\vspace{-2mm}
\section{BiasEye}
\par
\begin{figure*}[h]
    \centering
    \vspace{-4mm}
    \includegraphics[width=0.95\linewidth]{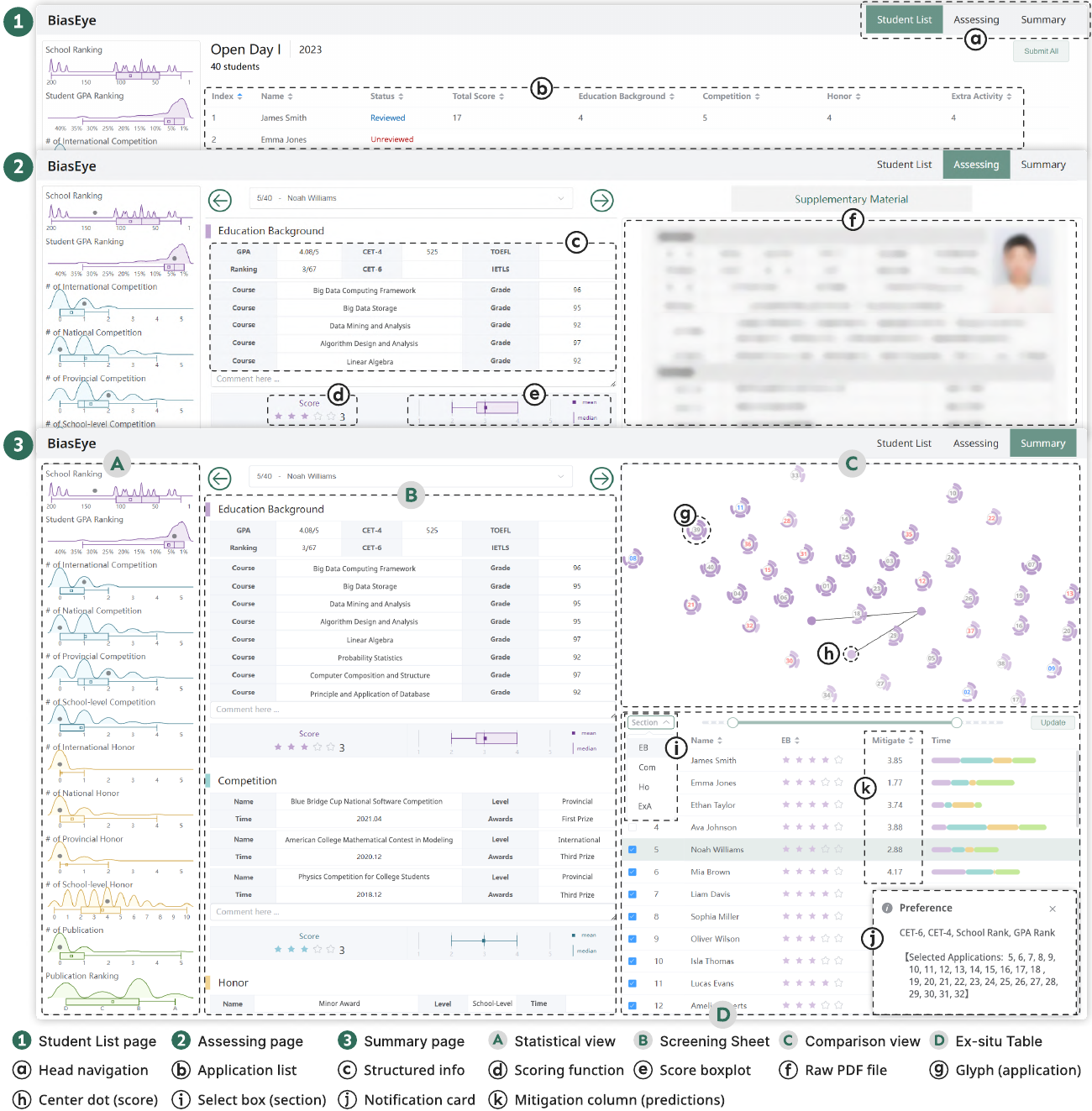}
    \vspace{-3mm}
    \caption{The front-end design of \textit{BiasEye}.}
    \label{fig:frontEnd}
     \vspace{-1mm}
\end{figure*}

 
In line with design goals \textbf{$\mathcal D1$}$\sim$\textbf{$\mathcal D5$}, we present \textit{BiasEye}, a real-time interactive system aimed at assisting reviewers in preventing, discovering, locating, and mitigating inconsistent decision-making. Implemented using Flask and Vue.js frameworks, it leverages Element-plus components\footnote{https://element-plus.org/} and D3.js\footnote{https://d3js.org/}~\cite{Bostock:2011:D3} for visualization. \textit{BiasEye} consists of three pages: 1) \textit{Student List}, displaying assigned applications and screening progress; 2) \textit{Assessing}, showing extracted information and original PDF materials for application assessment; 3) \textit{Summary}, offering retrospective bias-aware score inspection and revision through the \textit{Screening Sheet}, \textit{Comparison} view, and \textit{Ex-situ Table}. All three pages share a \textit{Statistical} view accessible via header navigation (\autoref{fig:frontEnd}-a). Potential usage methods and scenarios are explored in \autoref{sec:UsagePattern} to address inconsistent outcomes.

\vspace{-2mm}
\subsection{Screening Sheet}
\par Aligning with the admission committee's criteria, the \textit{Screening Sheet} includes a \textit{Basic Information} section and several screening sections, each with a unique color. In addition to the score\footnote{While each section employs a consistent score range of 1-5 points, the system assigns different weights to each section when computing the total score, as per the admission committee's guidelines.} \raisebox{-.38\height}{\includegraphics[height=3ex]{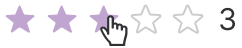}}
and the comment
\raisebox{-.32\height}{\includegraphics[height=2.9ex]{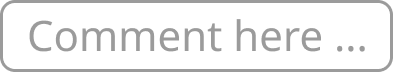}}
component involved in the original decision sheet mentioned in \textit{Finding 6}, each section also showcases structured entries extracted from resumes and included a box plot
\raisebox{-.38\height}{\includegraphics[height=4ex]{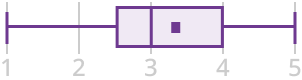}} 
showing statistical data for the assigned scores.

\par
\begin{figure*}[h]
\vspace{-2mm}
    \centering
    \includegraphics[width=0.92\linewidth]{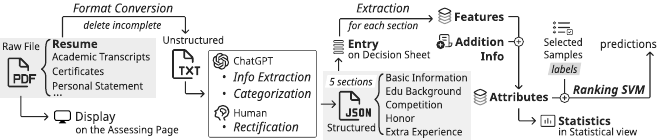}
    \vspace{-2mm}
    \caption{The overview of data processing and backend model pipeline of \textit{BiasEye}.}
    \label{fig:backEnd}
    \vspace{-3mm}
\end{figure*}

 As depicted in \autoref{fig:backEnd}, we first convert PDF files into TXT format and filter out resumes with incomplete or inaccurate information, ensuring the quality of our analysis. To extract information, we explored models and tools like \textit{CNN-BiLSTM-CRF}~\cite{Li:2021:AMRIEUBBC} and \textit{pyresparser}\footnote{https://github.com/OmkarPathak/pyresparser}, but these had suboptimal performance due to diverse resume formats and limited training samples. Consequently, we fine-tuned \textit{ChatGPT-4}\footnote{https://chat.openai.com}, implementing error correction codes and human verification for precision and consistency. Despite limitations like incomplete extraction of low-probability information with limited training data, this tool was effective in resume information extraction. Finally, as depicted in \autoref{tab:entry}, all raw text was extracted and structured into JSON format, encompassing five sections: \textit{Basic Information}, \textit{\EB{Educational Background}}, \textit{\Com{Competition}}, \textit{\Ho{Honors}}, and \textit{\ExA{Extra Activity}}. Letters of Recommendation (LoR) and Personal Statements (PS) are not displayed in this sheet, but users can access these original files directly from the \textit{Assessing} page.

\par Considering \textbf{$\mathcal D1$} in the \textit{PREVENTING} step and \textbf{$\mathcal D4$} in the \textit{LOCATING} step, we incorporate the \textit{Screening Sheet} with user-friendly interactivity into the \textit{Summary} page. This allows reviewers quick access to concise information about the selected applicant. While some details may be missing compared to the original PDF, the sheet provides ample cues. If more information is needed, reviewers can switch to the \textit{Assessing} page to examine the PDF. 

\begin{table*}[htbp]
    \vspace{-2mm}
    \begin{tabular}{ll}
        \toprule
        \textbf{Section}              & \textbf{Entries}                                                         \\ \hline
        \textbf{Basic Information}    & Name, Gender, Hometown, School, Major, Skill                             \\ \hline
        \textbf{Education Background} & GPA, Student Rank, CET-4, CET-6, TOFEL, IELTS, Course Name, Course Grade \\ \hline
        \textbf{Competition(*)}       & Name, Time, Level, Award                                                 \\ \hline
        \textbf{Honor(*)}             & Name, Time, Level                                                        \\ \hline
                                      & Project: Name, Time, Role, Description \\
        \textbf{Extra Activity(*)}    & Research Paper: Title, Author (order), Publication, Level, Summary \\
                                      & Other Experience: Name, Time  \\
        \bottomrule
    \end{tabular}
    \caption{Structured information entries from resumes in JSON formats, the extra activity section are divided into three sub-categories. *: The corresponding entries represent the content of each record in that section.}
    \label{tab:entry}
    \vspace{-4mm}
\end{table*}
\begin{table*}[h]
    \begin{tabularx}{\textwidth}{lX|p{14mm}}
        \toprule
        \textbf{Section}              & \multicolumn{2}{l}{\textbf{Attributes}}   \\
        \hline
        \textbf{Education Background} & CET-4, CET-6, TOFEL, IELTS  & \\
        \cline{1-2}
        \textbf{Competition ($\#$)}     & School Award, Provincial Award, National Award, International Award,  Mathematics Competition, English Competition, Computer Competition, Chemistry Competition, Electronics Competition, Mechanical Competition, Physics Competition, Biology Competition, Innovation and Entrepreneurship Competition, Other Competition &  \multirow{2}{*}{School Rank} \\
        \cline{1-2}
        \textbf{Honor ($\#$)}           & School Honor, Provincial Honor, National Honor, International Honor, Scholarship, Excellent Student, Outstanding Student, Outstanding Graduate, Student Officer, Volunteer, Social Practice, Skill Certificate & \multirow{2}{*}{Student Rank*}\\ 
        \cline{1-2}
        \textbf{Extra Activity ($\#$)}  & A-tier Publication, B-tier Publication, C-tier Publication, D-tier Publication, Projects, Project Manager, Project Participant & \\
        \bottomrule
    \end{tabularx}
    \caption{Attributes for each screening section, four sections share the attributes of School Rank and Student Rank. $\#$: The corresponding attributes represent the quantitative outcomes following aggregation. *: Indicates that the attribute has been normalized.}
    \label{tab:attr}
    \vspace{-4mm}
\end{table*}

\subsection{Statistical View}
\par
\begin{figure}[h]
    \centering 
    \vspace{-2mm}
    \includegraphics[width=\columnwidth]{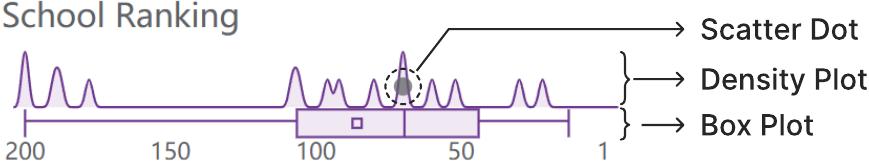}
    \vspace{-6mm}
    \caption{A visualized indicator of the \textit{Statistical} view.}
    \label{fig:statCard}
    \vspace{-3mm}
\end{figure}

 In response to \textbf{$\mathcal D1$}, we design the \textit{Statistical} view (\autoref{fig:frontEnd}-A) on \textit{BiasEye}'s left side. This view presents global statistics for the current application group, visualizing $12$ key indicators. These include school and normalized student GPA rankings, competition and honor count at various levels, and publication counts with corresponding conference/journal levels. Each indicator (\autoref{fig:statCard}) uses box plots to convey central tendencies and data dispersion, density plots to offer detailed distributional insights, and scatter dots to depict the cases of the currently selected students, offering an overarching perspective that aids \textit{PREVENTING} recency and contrast bias.

\par For each section, we defined a set of significant attributes denoted as $A={a_1, a_2, ..., a_M}$ based on feedback obtained during the formative study. These attributes serve as straightforward proxies for human decision-making preferences, as outlined in \autoref{tab:attr}. On one hand, most of these attributes are derived from features extracted directly from entries within the JSON file. Numerical and quantitative features, such as the count of different competition levels, can be readily obtained from the respective entries. Some features necessitate a text classification step before quantitative calculations, such as determining whether an applicant served as a manager or participant in a project. To facilitate this, we presented input and output samples to \textit{ChatGPT} and guided it through the classification process, providing explanations along the way. This approach aimed to encourage \textit{ChatGPT} to engage in a more deliberate thought process. On the other hand, two attributes, namely school ranking and publication ranking, were derived from additional information. This addresses \textbf{$\mathcal D1$} and aims to streamline the information search process, ultimately reducing the reviewer's workload. The school ranking is assigned a label from 1 to 200\footnote{University rankings: https://research.com/university-rankings/best-global-universities}, while the conference/journal level is categorized from A to D\footnote{Conference/journal level: https://research.com/}, where `D' signifies `unknown'.

\subsection{Ex-situ Table}

\begin{figure}[h]
    \centering 
    \vspace{-3mm}
    \includegraphics[width=0.7\columnwidth]{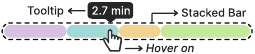}
    \vspace{-2mm}
    \caption{A stacked time bar \textit{Ex-situ Table}.}
    \label{fig:ExTimebar}
    \vspace{-4mm}
\end{figure}

\par As an extra enhanced version of the \textit{Student List} table, \textit{Ex-situ Table} incorporates additional visualizations and interactive features to address \textbf{$\mathcal D2$}, \textbf{$\mathcal D4$} and \textbf{$\mathcal D5$}. It provides an overview and facilitates score modifications, displaying application ID, applicant name, and section duration, which are calculated from time difference between two consecutive scoring events. Hovering over a  stacked bar (\autoref{fig:ExTimebar}) in `Time’ column reveals specific time values. Clicking a row in the table updates the \textit{Screening Sheet} and highlights the corresponding application glyph in the \textit{Comparison} view. The table dynamically displays the corresponding section column based on the selection in
\raisebox{-.32\height}{\includegraphics[height=2.9ex]{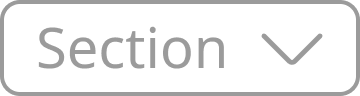}}, enabling direct score modification for \textbf{$\mathcal D5$}. Additionally, the \textit{Ex-situ Table} offers an interface that employs a machine learning method, specifically, \textit{Ranking SVM}, to help users \textit{DISCOVER} (\textbf{$\mathcal D2$}) inconsistent decision outcomes for each screening section. Through the use of a slider
\raisebox{-.38\height}{\includegraphics[height=3ex]{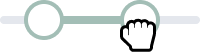}}
and checkboxes \faCheckSquareO, users can select a specific number of assessed applications as trusted training samples. Clicking the
\raisebox{-.32\height}{\includegraphics[height=2.9ex]{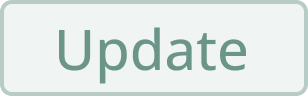}}
button activates the \textit{Ranking SVM} for analysis.


\par Inspired by \textit{Podium}~\cite{Wall:2018:Podium}, we employed \textit{Ranking SVM}~\cite{Klinkenberg:2000:DCDSVM} to automatically infer attribute weights from user-assigned application scores. This approach serves two purposes: Firstly, it helps reviewers examine their individual screening priorities and preferences, providing insight into how personal biases and emphases may affect their assessments. Secondly, \textit{Ranking SVM} forecasts future review tendencies using past records, indicating potential biases. Its low computational cost enables real-time monitoring, allowing reviewers to make timely adjustments and evaluate the appropriateness of modifications.

\par \textbf{Derive constraints.} \textit{Ranking SVM} optimizes a loss function involving pairwise constraints based on the Support Vector Machines (SVM) framework. We constructed a training set for the \textit{Ranking SVM} model using a subset of $k(>6)$ user-selected assessed applications, each assigned a score represented as $S$. We form pairs of data points $(d_i, d_j)$ with a label $l$. If $s(d_i)<s(d_j)$, we set $l=1$; otherwise, $l=-1$. For all pairs $i, j \in {1,\ldots, k}$ where $i \neq j$, we generated constraint tuples based on this criterion and treat all constraints as soft constraints.

\par \textbf{Calculate the ranking and transfer to score.} After training, we obtained a weight vector for the attributes to rank all the data items. We computed individual dot products of the weight vector ($w$) with each data item ($d_i$), resulting in an intermediate variable denoted as 
$v(d_i)=w \cdot d_i = \sum^{M}_{m=1}{w_m \cdot a_m}$, 
where $a_m$ represents the attribute value in the selected section. Subsequently, we mapped the values of $v$ to the interval $[min(S)-0.5, max(S)+0.5]$, preserving two decimal places to enhance transparency and facilitate explanation. This mapping yielded the prediction score $S^{\prime}$, with the condition that ${s^{\prime}}_i=0$ if $s_i=0$.

\par The prediction score is displayed in the `Mitigate' column, and a notification appears in the bottom right of the page, listing the top $k(=10)$ significant model attributes and training application IDs. Users can \textit{LOCATE} (\textbf{$\mathcal D4$}) inconsistent decision outcomes by comprehensively comparing these predictions with their scores and cross-referencing this information with the significant attributes and original data.


\subsection{Comparison View}

\begin{figure}[h]
    \centering 
    \vspace{-3mm}
    \includegraphics[width=\linewidth]{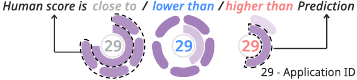}
    \vspace{-6mm}
    \caption{Design of glpyhs in \textit{Comparison} view.}
    \label{fig:Glyphs}
    \vspace{-3mm}
\end{figure}

\par To address \textbf{$\mathcal D1$}, we developed a visual glyph (\autoref{fig:Glyphs}) for comparing human scores and model predictions. Each glyph corresponds to an applicant, with the number denoting the application ID, the outer ring encoding the human score and the inner ring encoding the prediction. A linear color scheme is used for both rings, facilitating the rapid identification of applications with varying scores. The ID color indicates whether the human score is \higherthan{higher}/\lowerthan{lower} than or \closeto{close to} the prediction, highlighting inconsistencies in human scores and their direction. To \textit{LOCATE} (\textbf{$\mathcal D4$}) anomalies among similar applications, glyph position is determined using the \textit{t-SNE}~\cite{van:2008:visualizing} method based on the attributes of selected section, ensuring similar applications are closer together. Solid dots represent high-dimensional centers of all applicants who received the same human score, follow the same color scheme as the glyph rings and are connected from lowest to highest scores (\textbf{$\mathcal D2$}). Furthermore, to provide visual aid for \textbf{$\mathcal D4$}, hovering over a glyph or center highlights applicants with the same human score.

\section{User Study} \label{sec:userStudy}

\par Obtaining institutional IRB approval, we conducted a user study involving $20$ participants with mixed backgrounds. The primary aim of this study was to assess the effectiveness of our bias-aware design. To achieve this, we aimed to address three key research questions (\textbf{$\mathcal RQ1-RQ3$}) through our evaluation process.
\begin{itemize}
    \item \textbf{$\mathcal RQ1$}: How are the usability and effectiveness of the bias-aware system in material screening?
    \item \textbf{$\mathcal RQ2$}: How will participants interact with and be affected by the bias-aware system in material screening?
    \item \textbf{$\mathcal RQ3$}: How will participants trust and collaborate with the ML method?
\end{itemize}

\subsection{Experiment Setup} \label{sec:ExpSetup}
\subsubsection{Dataset}
\par We obtained IRB approval for data collection and used a dataset from a local university's information science master's program. Graduate admission, similar to college admission, emphasizes merit and alignment with the institution's mission. We randomly selected two groups of 40 complete applications (excluding incomplete ones) for a preliminary trial and a formal experiment. Each application included a resume, academic transcripts, a personal statement (PS), and up to two letters of recommendation (LoR), all in PDF format. To ensure the experiment's completion within 1.5 hours, we retained only the resume, transcripts, and certificates. Notably, these materials were from past admission interviews, and we had only raw PDF files, making it impossible to verify results with reliable ground truth. Additionally, we anonymized identifiable details like names and photos.

\subsubsection{Baseline System and Control Conditions}
\par We adopted a two-pronged approach to assess the effectiveness of our system. First, we used a \textbf{between-subject design to evaluate the \textit{Statistical} view}, dividing participants into two groups randomly: Group A used the baseline system, and Group B used \textit{BiasEye}. Both systems consisted of three pages (\autoref{fig:frontEnd}-~\darkcircled{1}~\darkcircled{2}~\darkcircled{3}), but the baseline system lacked the \textit{Statistical} view and publication level in the \textit{Screening Sheet} (\autoref{appendix:baselineSystem}). Both systems were hosted on a web server, accessible to participants via public links. Second, we used a \textbf{within-subject design to evaluate the \textit{Summary} page} in two stages. In stage I, participants could only use the \textit{Student List} and \textit{Assessing} pages. In stage II, participants could further adjust their decisions using the entire system.

    \begin{figure*}[h]
    \vspace{-5mm}
        \centering
        \includegraphics[width=\linewidth]{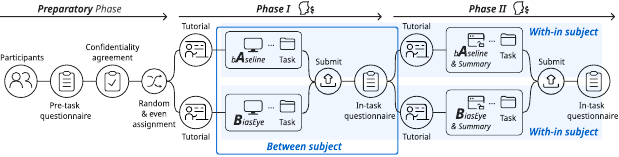}
        \vspace{-7mm}
        \caption{Procedure of user study.}
        \label{fig:procedure}
          \vspace{-2mm}
    \end{figure*}
    
\begin{table*} [ht]
\footnotesize
\centering
\begin{tabular}{c c c c c | c c c c c}
\hline
ID & Gender/Age & Degree & Experienced & Condition Group & ID & Gender/Age & Degree & Experienced & Condition Group \\ \hline
P1 & Male / 21   & Bachelor & N & A & P2 & Male / 23   & master & Y & B \\
P3 & Male / 24   & master   & N & A & P4 & Male / 24   & master & N & B \\
P5 & Female / 23 & master   & Y & A & P6 & Female / 24 & master & Y & B \\
P7 & Male / 23   & master   & Y & A & P8 & Male / 23   & master & Y & B \\
P9 & Male / 23   & master   & Y & A & P10 & Male / 21  & master & N & B \\
P11 & Male / 21  & Bachelor & Y & A & P12 & Male / 25  & Ph.D & Y & B \\
P13 & Female / 26 & Ph.D    & N & A & P14 & Female / 24 & Bachelor & N & B \\
P15 & Male / 22  & master   & Y & A & P16 & Male / 21  & Bachelor & Y & B \\
P17 & Male / 21  & Bachelor & N & A & P18 & Female / 23 & master & N & B \\
P19 & Female / 22 & master  & N & A & P20 & Male / 21  & Bachelor & N & B \\ \hline

\end{tabular}
\caption{Demographic information of participants. Experienced means one has prior involvement in relevant screening assistance scenarios encompassing over 20 applications. Group A uses Baseline system, group B uses \textit{BiasEye} system.}
\label{tab:demographics}
\vspace{-7mm}
\end{table*}

\subsection{Participants}
\par We recruited 20 participants (P1 to P20): 14 males and 6 females. Among them, 6 held bachelor's degrees, 12 held master's degrees, and 2 held Ph.Ds. Participants were evenly divided into the experiment (B) and control (A) groups based on demographics (\autoref{tab:demographics}). Before the formal experiment, all participants signed a confidentiality agreement, became familiar with the training program and department's mission. Special attention was given to those without prior relevant experience (n > 20) to ensure they understood the screening expectations. Their participation was incentivized by performance-based compensation.

\subsection{Task and Procedure}
\subsubsection{Task}
\par We simulated a real-world material screening scenario for user study. Participants were instructed to act as students in a Human-Computer Interaction (HCI) laboratory, tasked with preliminarily review $40$ admission applications due to the time constraints of their professor. The participants' responsibility was to consider multiple factors like personal backgrounds, experiences, abilities, and the lab’s requirements. Their anonymous screening outcomes would be combined with others to determine final screening results. To fulfill this task, participants had to: 1) assign scores to each application in four sections: \textit{\EB{Education Background (EB)}}, \textit{\Com{Competition (Com)}}, \textit{\Ho{Honor (Ho)}} and \textit{\ExA{Extra Activity (ExA)}}, which were chosen based on the actual department criteria. 2) They were prohibited from discussion and communication, and 3) were not required to consider score weighting within each section. 4) They were encouraged but not forced to aim for an average score of $3$ in each section. Additionally, online references\footnote{https://research.com/} including school rankings, conference and journal rankings, and a formal document listing the level of college student competitions were provided for assistance.

\subsubsection{Procedure}


\par \autoref{fig:procedure} outlines our mixed-subject experiment. Before the study, participants signed confidentiality and completed a pre-task questionnaire collecting demographics. We introduced the experimental task and its objectives in a comprehensive manner, \textbf{rather than explicitly disclosing the focus on cognitive bias}, we emphasized the core principle of \textit{individual fairness} and underscoring the gravity of inconsistent outcomes. This approach ensured that participants remained unaware of the precise nature of our study.

\par Next, we introduced the system corresponding to their belonging condition in stage I and provided a set of toy trial materials for familiarization. During Phase I, participants were allotted $50$-$70$ minutes to complete the task as consistently as possible, then submitted their results and filled out an in-task questionnaire. The main goal of Phase I is to assess how the introduction of the \textit{Statistical} view impacts the consistency of participants in decision-making. To address potential residual effects between the two experiments and reduce response bias within the two conditions, we adopted a between-subject design approach.

\par Moving to Phase II, we introduced the \textit{Summary} page and the \textit{Ranking SVM} model, which learns participants' screening preferences and predicts scores. To directly compare the change in decision-making before and after model intervention, we utilized a within-subject design independently for both groups. Simultaneously, both groups maintained a between-subject design that included the \textit{Statistical} view as a variable. Participants were given $20$ minutes to revise their outcomes with the assistance of \textit{Summary} page. Subsequently, they submitted again and completed a post-task questionnaire. 

\par Two of the authors  acted as experimenters to ensure smooth progress and provided assistance as needed. The study spanned approximately two hours, with participants receiving USD $12$ compensation on average.

\vspace{-2mm}
\subsection{Data Collection}

\par We conducted a general quality check for each participants by examining the usage time of Phase I, which started when they began the task and ended at their first outcome submission.  One submission from group B (P20) was rejected due to a extraordinarily short duration ($30$ minutes) for Phase I. Besides, one scoring log files form group A (P1) were irreversibly corrupted, we excluded his log files and questionnaires from quantitative analysis but kept video for qualitative analysis. We ended up with 18 valid responses, 9 per group.  All data will be used solely for experimental outcome analysis and won’t be shared or disclosed non-anonymously.

\vspace{-1mm}
\subsection{Measurement}
\par For both the in-task and post-task questionnaire, we utilized a $7$-point Likert scale (1: Not at all/Strongly disagree, 7: Very much/Strongly agree, and a $10$-point scale for workload-related questions) to collect participants' feedback on the respective systems and their attitudes toward their own results in different phases of the study. First, in line with the \textbf{System Usability Scale} (SUS)~\cite{Brooke:2013:SUS}, we crafted questions primarily including: 1) Ease of use; 2) Ease of learn; 3) System satisfaction; and 4) Likelihood of future use; Second, in terms of \textbf{Self-Evaluation}, we designed questions mainly including: 1) Consistence criteria; 2) Degree of distinction; 3) Fewer revisions; and 4) Perceived efficiency promotion. Third, drawing from the NASA-TLX survey~\cite{Hart:1988:Development}, we posed questions about \textbf{Workload Assessment}, including: 1) Psychological workload; 2) Physical workload; 3) Time workload; and 4) Level of frustration. Fourth, as for \textbf{System Design}, we tailored questions concerning the \textit{Statistical} view for group B participants in the in-task questionnaire and regarding the \textit{Ex-situ Table} and \textit{Comparison} view for both groups in the post-task questionnaire, including: 1) Intuitive visualization; 2) Convenience of interaction; and 3) Overall helpfulness. Additionally, we also included optional subjective questions for qualitative insights. Participants were instructed to ``think aloud’’ throughout while their screens and audio were recorded. The system documented the section name and score for every modification in scoring logs during both phases for later quantitative analysis in \autoref{sec:studyResults}.

\section{Results and Analysis} \label{sec:studyResults}
\par This section organizes quantitative and qualitative results for research questions $\mathcal RQ1$ to $\mathcal RQ3$. Our \textbf{quantitative analysis}, besides descriptive statistics, employed the Mann-Whitney U test~\cite{Mann:1947:Test} to investigate differences between groups using different systems, and the Wilcoxon signed-rank tests~\cite{Wilcoxon:1970:Critical} to evaluate disparities between groups of participants using the same system. For our \textbf{qualitative analysis}, one author transcribed participants’ screen recordings, capturing system usage and reactions to potential inconsistent decisions. Two authors then coded these transcriptions using thematic analysis~\cite{Guest2011:ThematicAnalysis}, with specific examples included in this paper.

\subsection{RQ1. How are the usability and effectiveness of the bias-aware system in material screening?}

    \begin{figure}[h]
    \centering 
    \vspace{-3mm}
    \includegraphics[width=\linewidth]{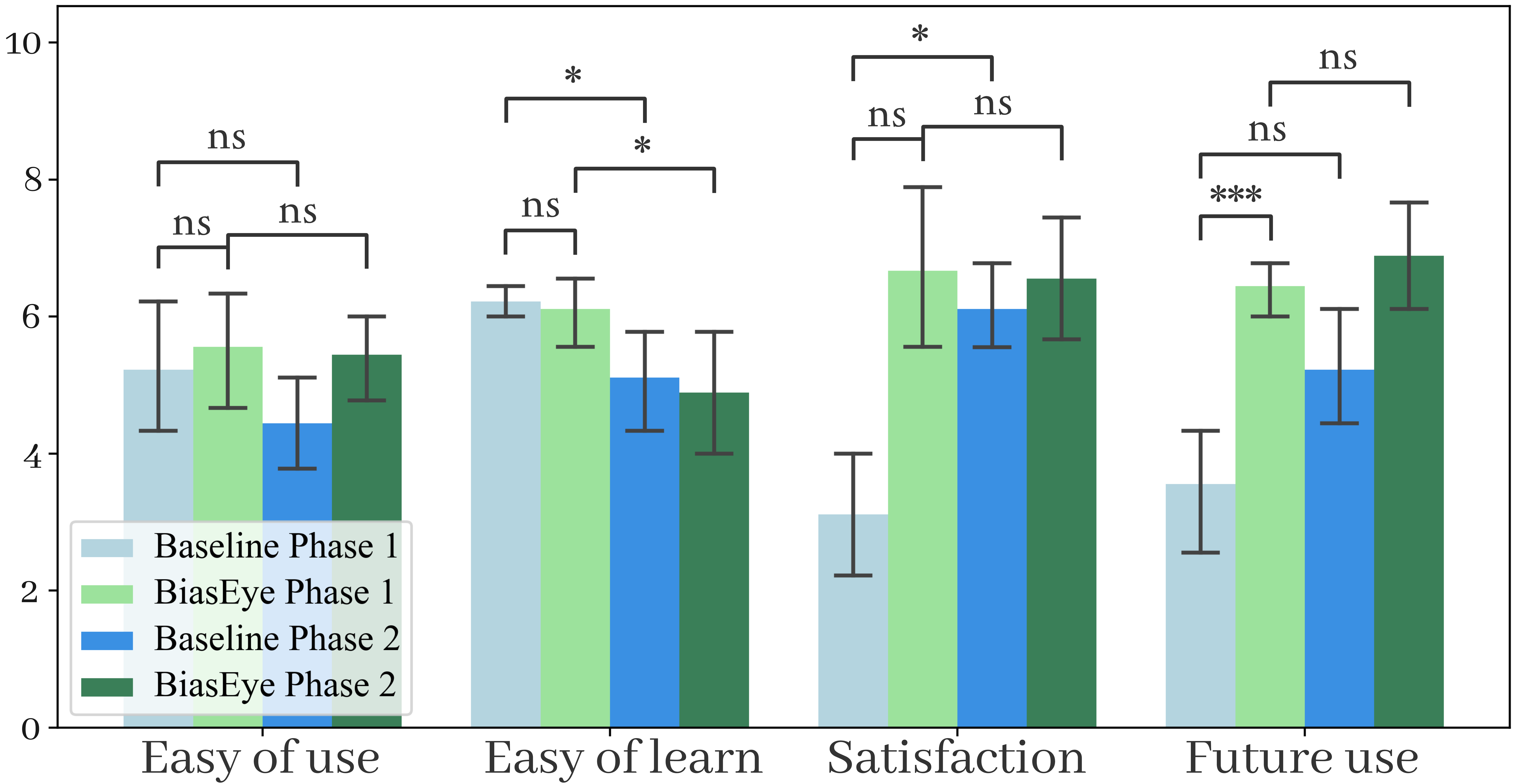}
    \vspace{-6mm}
    \caption{The usability of usefulness of the system. Error bars indicate standard errors. (ns: p < .1; $^{\ast}$: p < .05; $^{\ast\ast}$: p < .01; $^{\ast\ast\ast}$: p < .001).}
    \label{fig:usability}
    \vspace{-3mm}
\end{figure}


\par As shown in \autoref{fig:usability}, the questionnaire presents participant ratings of system usability at various stages and with different systems. When comparing the Phase 1 data for both systems, we observed that the \textit{BiasEye} system did not lead to a significant increase in `ease of use' or `ease of learning'. However, it did demonstrate a substantial increase in `satisfaction' ($U=3.5, p<0.01$) and `future use' ($U=1.5, P<0.001$).

\par Conducting a comparative analysis of data within the same system at different phases, we noticed that the introduction of the Summary Page had a significant impact. Specifically, it led to a decrease in `ease of learning' for both the Baseline and \textit{BiasEye} systems ($T=0.0, p < 0.05$ in Baseline, $T=0.0, p < 0.05$ in \textit{BiasEye}). Furthermore, it significantly enhanced `satisfaction' ($T=0.0, p < 0.05$) in the case of the Baseline system. However, there were no significant changes in terms of `ease of use' and `future use' for both systems.

\par Moving forward, we proceed to evaluate the efficacy of the \textit{BiasEye} system by delving into the data collected from participants as they engaged in the real scoring process. Our analysis has unveiled the following two key findings.


\par
\begin{figure}[h]
  \centering
  \vspace{-4mm}
  \begin{minipage}[t]{\columnwidth}
    \centering
    \label{fig:change_1}\includegraphics[width=0.92\columnwidth]{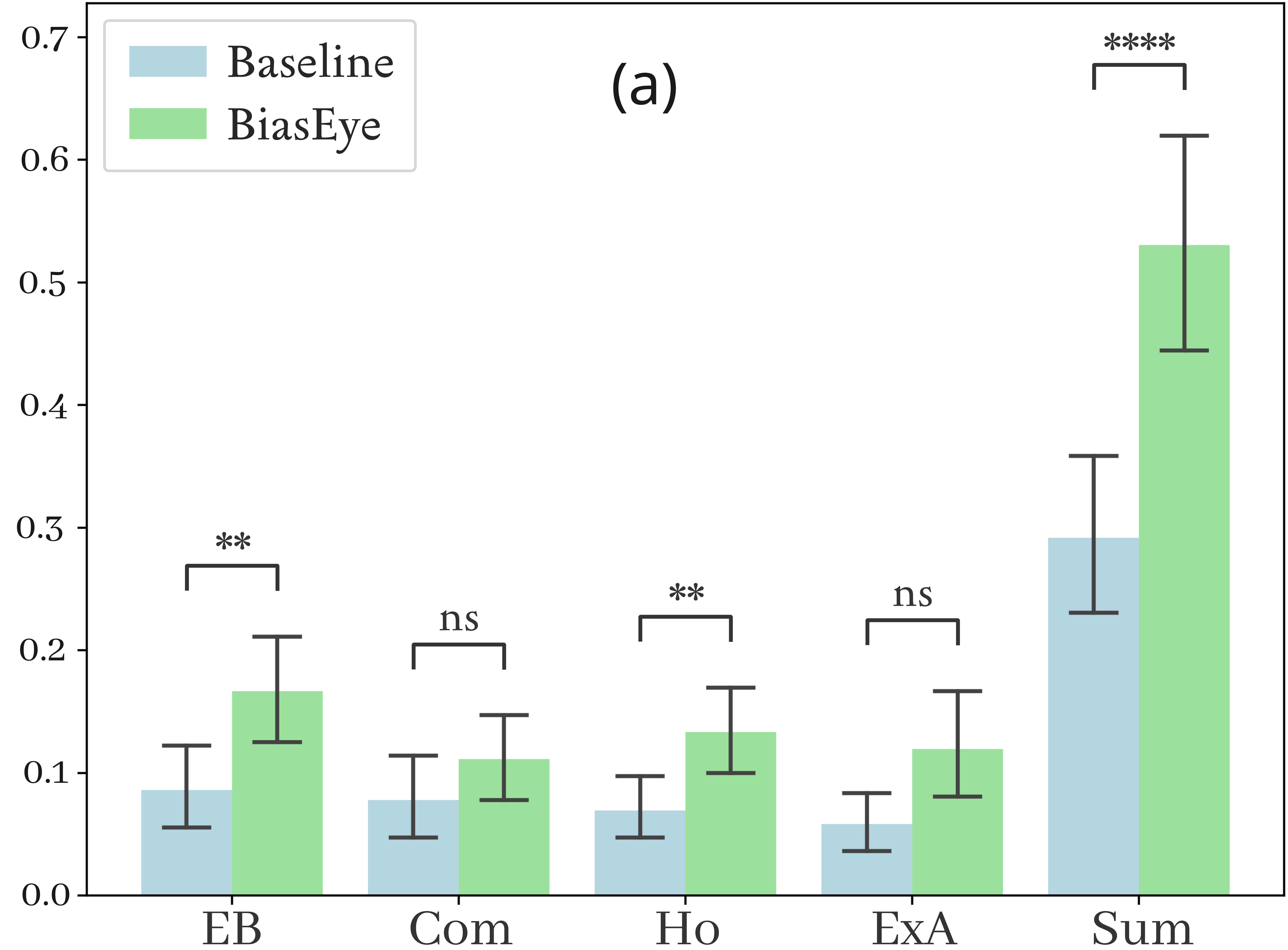}
  \end{minipage}
  \begin{minipage}[t]{\columnwidth}
      \centering
      \label{fig:change_2}\includegraphics[width=0.92\columnwidth]{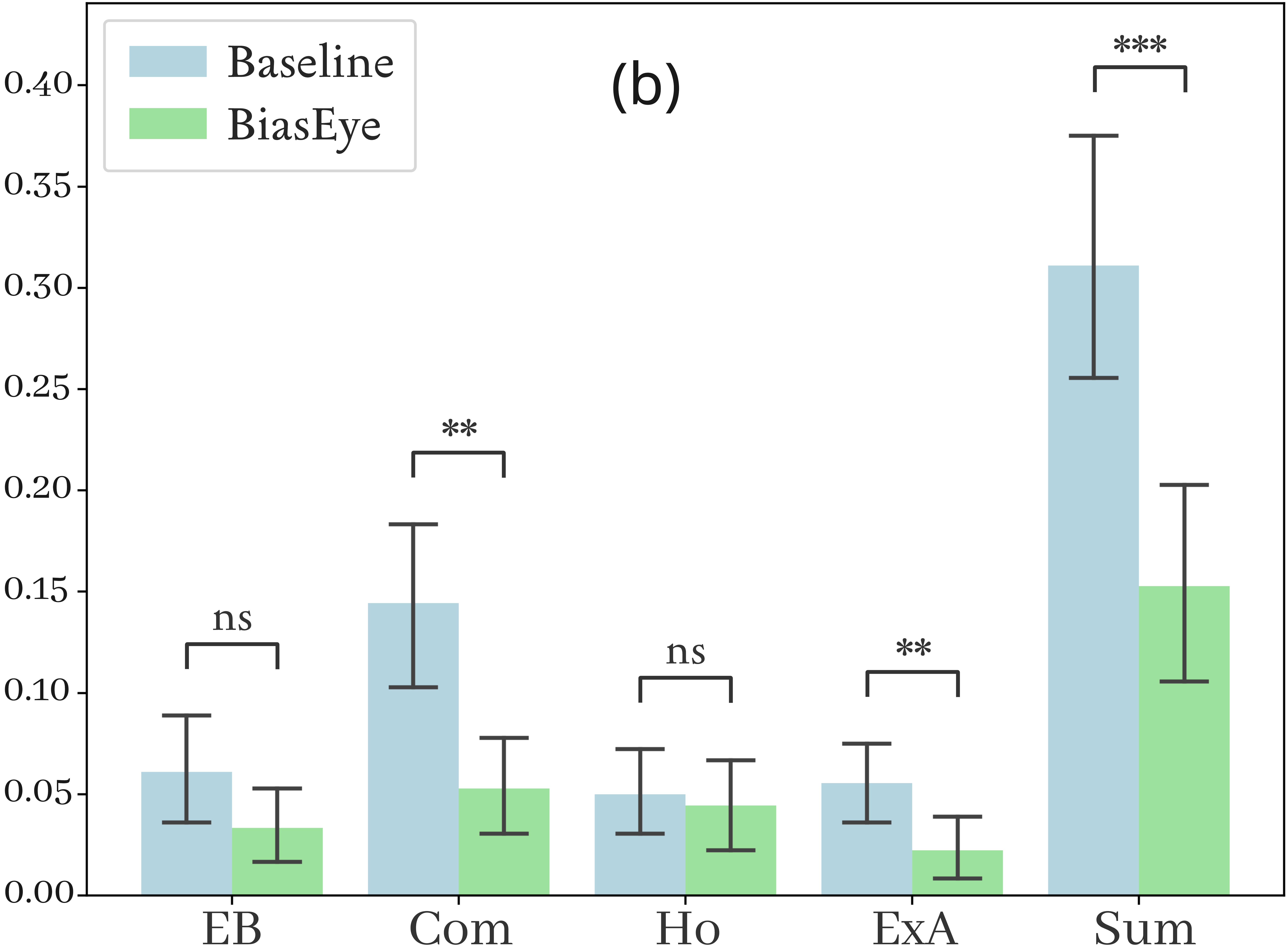}
  \end{minipage}
  \begin{minipage}[t]{\columnwidth}
      \centering
      \label{fig:change_time}\includegraphics[width=0.9\columnwidth]{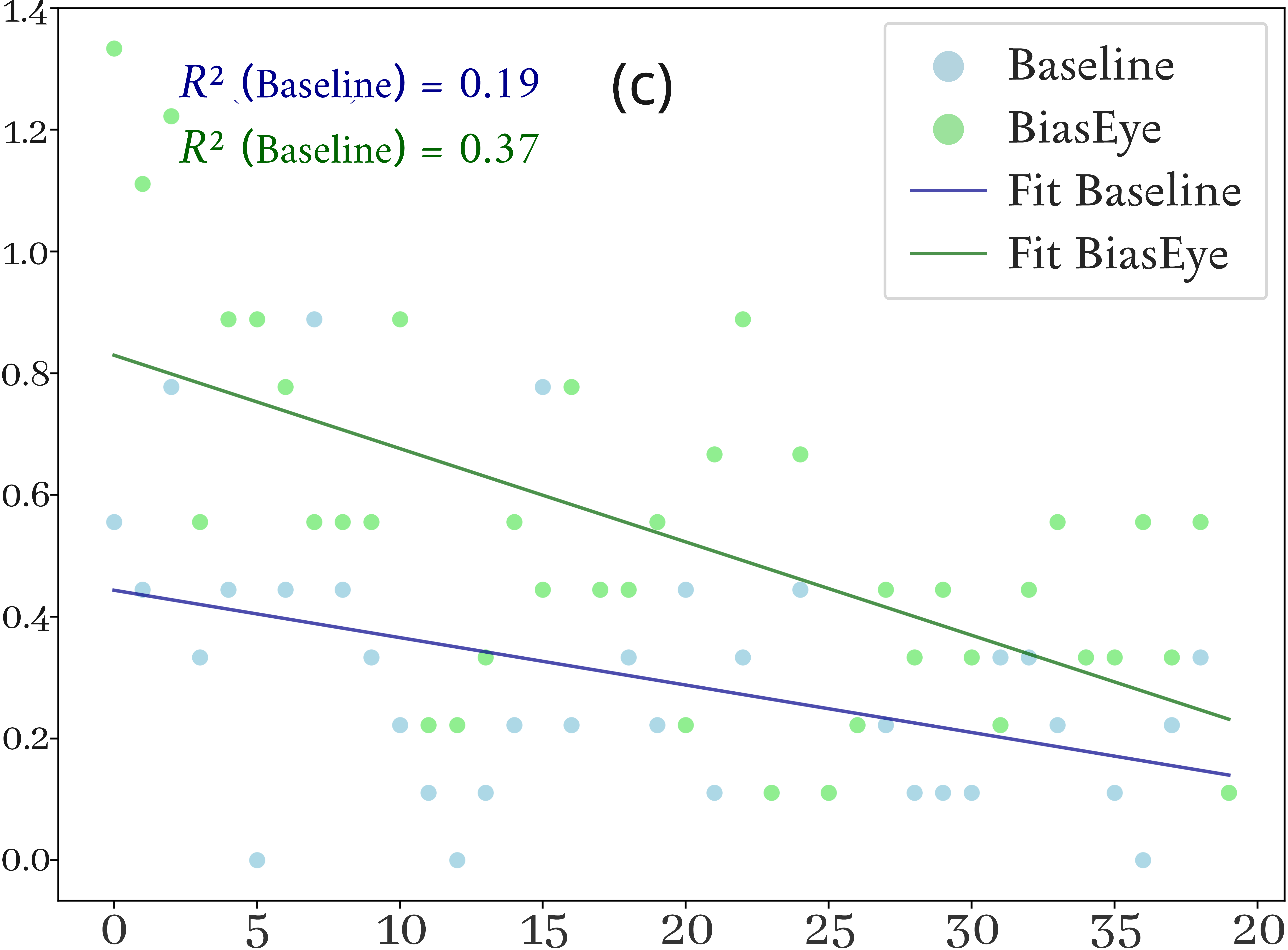}
  \end{minipage}
  \caption{Differences in Revision Behavior Among Participants in Different Groups. (a) Differences in revision behavior during Phase I. (b) Differences in revision behavior during Phase II. (c) The average number of score modifications varies throughout the screening process in Phase I, with the horizontal axis representing application ID. The error bars indicate standard errors. (ns: p < .1; $^{\ast}$: p < .05; $^{\ast\ast}$: p < .01; $^{\ast\ast\ast}$: p < .001; $^{\ast\ast\ast\ast}$: p < .0001)}
  \label{fig:change}
  \vspace{-4mm}
\end{figure}

 \textbf{Finding 7: The \textit{Statistical} view and additional information facilitates participants in raising awareness of bias in the process and proactively reducing inconsistencies in decision-making.} Our findings stem from an examination of participants' interactions with the system, focusing on instances where they adjusted their initially assigned scores. The statistical analysis of score revisions during Phase I and Phase II is presented in \autoref{fig:change}(a) and \autoref{fig:change}(c), respectively.

\par In \autoref{fig:change}(a), it becomes apparent that participants using the \textit{BiasEye} system displayed significantly higher average frequencies of score revisions for the \EB{EB} ($U=522, p < 0.01$), \Ho{Ho} ($U=539, p < 0.01$), and Sum ($U=389, p < 0.001$) categories compared to those using the Baseline system. As there was no machine learning intervention in Phase I, participants adjusted their decision outcomes relying on personal judgment. The transcripts indicate that participants recognized the inconsistency in their initial decisions, and their perception of this inconsistency became more pronounced and less ambiguous. Participants demonstrated the ability to discern candidates with varying qualifications more swiftly and accurately.

\par To reinforce this observation, we plotted a scatter plot in \autoref{fig:change}(c) depicting the average number of score changes against the applicant sequence, fitting a linear function. As depicted, as the number of students being scored increased, both groups experienced a decline in the frequency of revisions. This observation aligns with the expectation that participants' evaluation criteria improve and stabilize over time. Notably, the fitted line for \textit{BiasEye} users is always higher than the baseline, suggesting that the proposed system increased the number of revisions generally, rather than being influenced by outliers.

\par The results from Phase II further substantiate the notion that the \textit{BiasEye} system contributes to the decrease of inconsistent decisions. As illustrated in \autoref{fig:change}~(b), participants utilizing the \textit{BiasEye} system exhibited significantly lower frequencies of revisions in the \Com{Com} ($U=1,086, p < 0.01$), \ExA{ExA} ($U=1,028, p < 0.01$), and Sum ($U=1,181, p < 0.001$) categories. Despite being exposed to more comprehensive global information in Phase II, participants employing the \textit{BiasEye} system had already mitigate most of inconsistent decision outcomes during Phase I, decrease the requirement for additional score revisions. P10 explicitly stated, ``\textit{without those charts on the left (\textit{Statistical} View), it can be kind of hard for me to tell the difference between the different application levels because the scores start to blur together. Having those charts really makes a difference for me.}''

\par \textbf{Finding 8: Participants utilizing the \textit{BiasEye} system exhibit more concentrated scoring for high-quality applicants, resulting in fewer instances of inconsistent outcomes.} While cognitive bias can play a role, it's important to recognize that different reviewers may hold varying opinions about an application. Existing literature, as mentioned in Coleman et al.~\cite{coleman:2018:understanding}, emphasizes the use of ``interrater reliability'' to ensure the effectiveness and consistency of screening decisions. One way to assess this is through ``composite reliability'', as outlined by Coleman and colleagues~\cite{coleman:2018:understanding}, where a group of reviewers score within an acceptable range.

\par
\begin{figure*}[h]
\begin{centering}
\includegraphics[width=\linewidth]{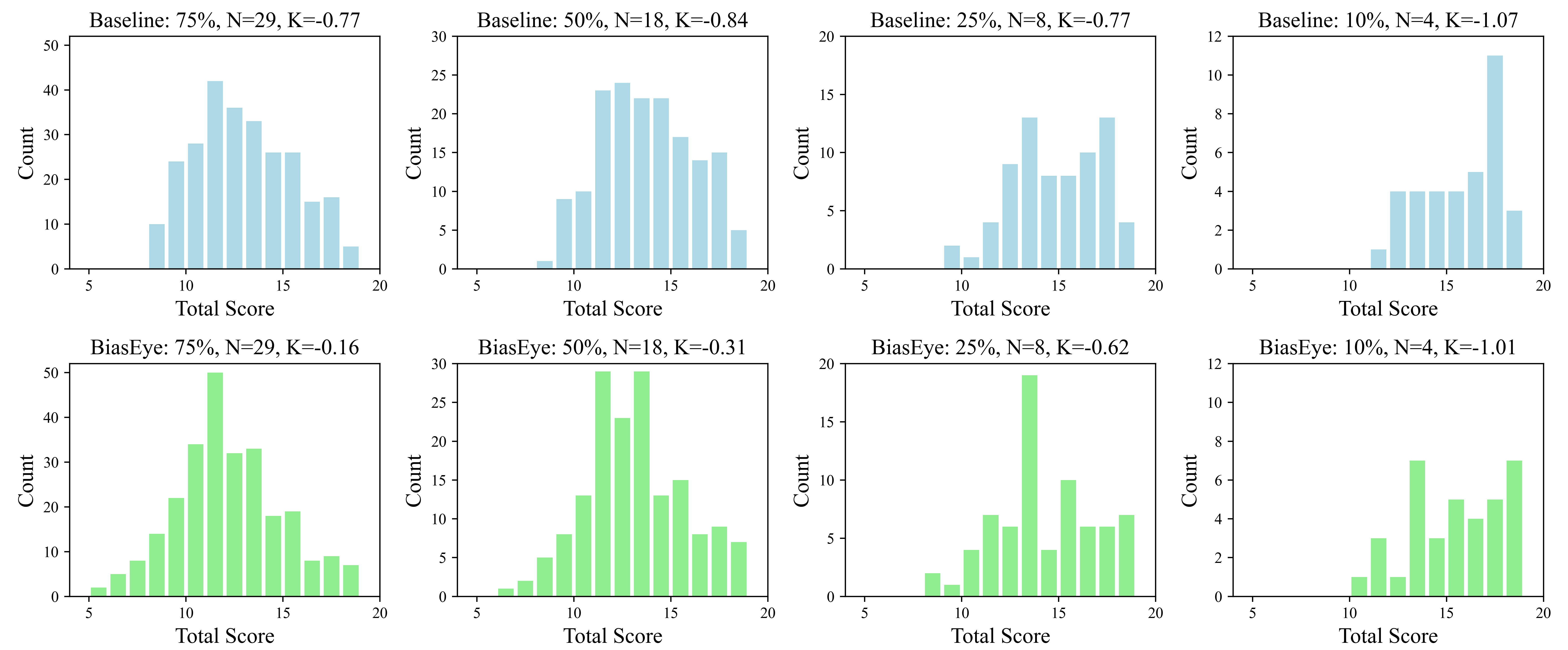}
\end{centering}
\vspace{-6mm}
\caption{Differences in the score distribution between Baseline and \textit{BiasEye} systems in Phase I.}
\label{fig:score_dist_p1}
    \vspace{-2mm}
\end{figure*}

 To evaluate whether the \textit{Statistical} view in the \textit{BiasEye} system helps mitigate screening inconsistencies, we compared the screening outcomes for high-quality applications in Phase I at different score levels (assuming equal section weights) in both Group A and Group B. We took the intersection of the results from both groups to ensure consistency. The outcomes are presented in \autoref{fig:score_dist_p1}, where each bar represents the number of applications receiving a specific score. Here, we denote the number of compared applications as $N$ and measure the kurtosis of the histograms as $K$.

\par
\begin{figure*}[h]
    \centering
    \includegraphics[width=\linewidth]{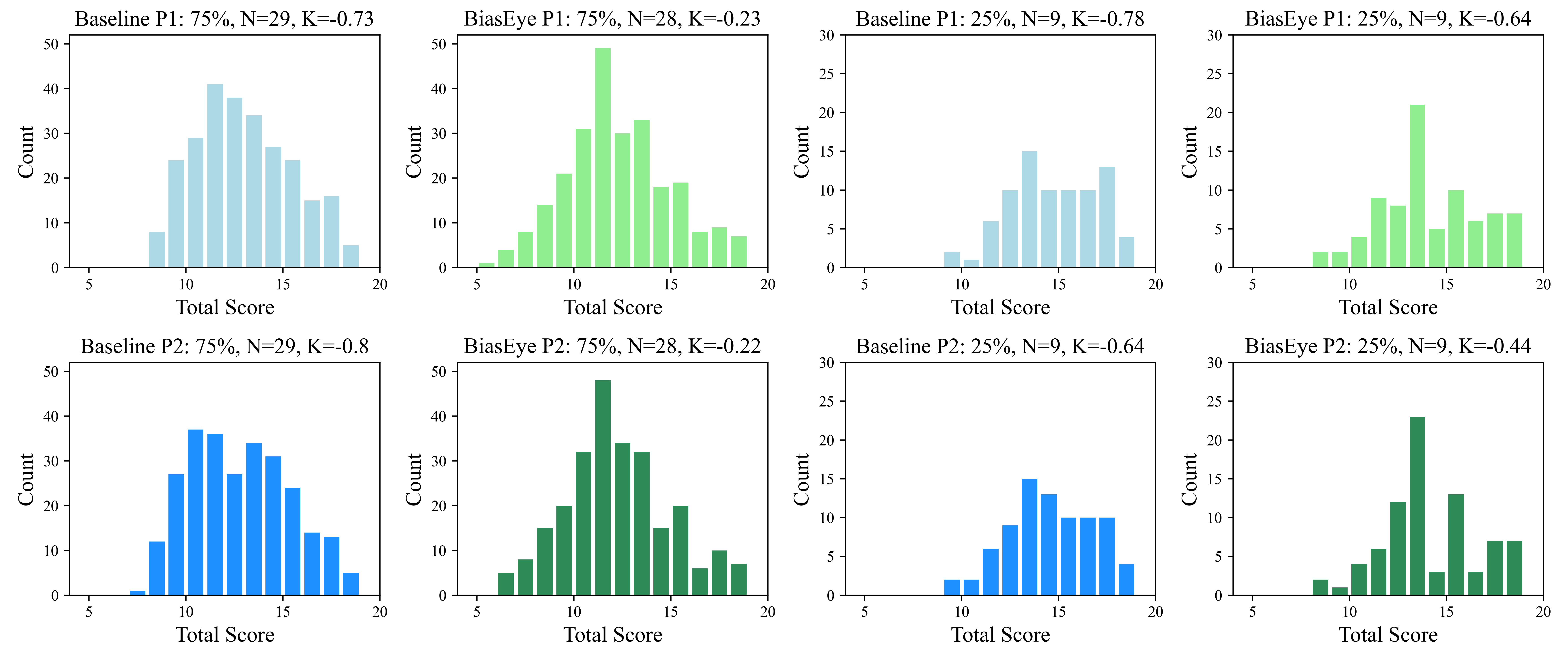}
    \vspace{-6mm}
    \caption{Differences in the score distribution between Phase I (P1) and Phase II (P2).}
    \label{fig:score_dist_p2}
    \vspace{-3mm}
\end{figure*}

 Our observations indicate that Group B, using the \textit{BiasEye} system with the \textit{Statistical} view, exhibits more centralized outcomes, reflected in the higher kurtosis value ($K$). A similar trend is observed when comparing Phase I to Phase II, regardless of whether the Baseline or \textit{BiasEye} system was used. This evaluation underscores the effectiveness of the \textit{Summary} page in mitigating inconsistencies, as seen in \autoref{fig:score_dist_p2}. It's worth noting that the phenomenon in \autoref{fig:score_dist_p2} is less pronounced due to the comparison being based on an intersection, which excludes a significant portion of adjusted applications in Phase II.

\vspace{-1mm}
\subsection{RQ2. How will participants interact with and be affected by the bias-aware system in material screening?}
\par Building upon the earlier-discussed analysis methods outlined in \autoref{sec:studyResults}, we initially explore the ways in which participants will engage with our bias-aware designs to fine-tune their decision outcomes. Subsequently, we proceed to unveil the discoveries regarding how these designs impact participants' cognitive workload and their self-evaluation of the decisions made.

\vspace{-2mm}
\subsubsection{Usage pattern.} \label{sec:UsagePattern}
\par This section presents the observations regarding how participants utilize our system design in a systematic four-step process aimed at preventing, discovering, locating, and mitigating inconsistent decision outcomes.

\par \textbf{Step 1. Preventing.} Video transcription shows all Group B participants employed the \textit{Statistical View} for insights into application materials and positioning applicants, and considered publication level as a screening criterion. Conversely, most Group A participants (7 out of 10) tried to check the given reference, but nearly all (6 out of 7) stopped after about 20 applications. Interestingly, three participants ignored this feature. These observations support our design motivation to enhance information transparency and accessibility.

\par \textbf{Step 2. Discovering.} Participants predominantly employ two categories of methods to identify inconsistencies in their decision outcomes on the \textit{Summary} page. First, a minority of participants (3 out of 19) inspected exceptions to the time allocation in the \textit{Ex-situ Table}. Second, the majority of participants (17 out of 19) used the back-end model to aid them in discovering potential anomalies through prediction scores. They selected trusted samples for back-end training through three distinct approaches:
\begin{itemize}
    \item Most participants (13 out of 19, including 7 from Group A) directly chose applicants falling within a specific range based on the screening order. This range deliberately excluded the initial $5$-$10$ applicants, as participants perceived their screening criteria to be either more lenient or stricter for this subset. This observation suggests that cognitive bias cannot be entirely eliminated, even with the aid of statistical information and supplementary.
    \item A subset of participants (5 out of 19, including 2 from Group A) manually selected representative applicants from each score category ($1$-$5$) using checkboxes.
    \item Participant P9 employed an unconventional approach that exceeded our expectations in sample selection. Initially, P9 included all applicants in the first round of training and then chose applicants exhibiting consistency between the model's predicted scores and human scores as the final samples for the second round of training. In the video, P9 mentioned being unfamiliar with machine learning but believed this approach could help identify samples that could serve as representatives of his scoring criteria. We observed an increase in the number of consistent outcomes after the second round of training, although this may have occurred by chance. Exploring whether repeating such operations could lead to convergence and automate the process is an intriguing topic for future research.
\end{itemize}

\par \textbf{Step 3. Locating.} Building upon the methods outlined in Step 2, participants employed specific strategies to identify anomalies. This process can be categorized into two distinct approaches. First, a minority of participants (3 out of 19) directly scrutinized applicants who received either inadequate or excessive time allocations, classifying them as cases of oversight or difficulty in decision-making, respectively. Second, participants who utilized the back-end model (comprising 17 out of 19) employed two primary methods to pinpoint applicants with potential inconsistencies:
\begin{itemize}
    \item Five participants harnessed the sorting function within the `Ex-situ Table'. They initially sorted the table based on the columns labeled `\EB{EB}/\Com{Com}/\Ho{Ho}/\ExA{ExA}' or `Mitigate'. Their focus was directed towards applicants where the order of predictions/human scores contradicted the ascending or descending order of human scores/predictions.
    \item Twelve participants identified potential inconsistencies by observing the ID color and assessing the variance between the two rings of a glyph. When confronted with multiple anomalies marked with \higherthan{blue} or \lowerthan{red} colors, participants developed distinct patterns of focus: i) A majority (7 out of 12) concentrated on applicants displaying a high discrepancy between the two rings, a preference influenced by their personal perception. ii) Two participants focused on identifying the \lowerthan{lower}/\higherthan{higher} scores within an overall trend of \higherthan{higher}/\lowerthan{lower} scores. iii) Three participants searched for inconsistencies within the pool of applicants who had received high human scores.
\end{itemize}
These patterns of focus shed light on participants' expectations of generating rational screening outcomes. 
            
\par \textbf{Step 4. Mitigating.} Participants accessed the \textit{Screening Sheet} of the corresponding student on the \textit{Summary} page by clicking on rows within the \textit{Ex-situ Table}. They employed various strategies to adjust the assigned scores, which included: (1) Comparing the applicant's score with those who received the same score; (2) Comparing the applicant's score with individuals who had similar predicted scores from the model; (3) Comparing the applicant's score with students positioned closely in the \textit{Comparison} view. (4) Relying entirely on, or taking into consideration, the model's recommendations; (5) Referring to the keywords listed in the notification card to understand the model's rationale and checking if any relevant features were overlooked during Phase I; (6) Assessing the model's performance based on keywords and the distribution of ID colors in the \textit{Comparison} View to determine whether further examination of potentially inconsistent applications was necessary. The sixth strategy is particularly relevant to the issue of trust in the model, and our findings related to this are presented in~\autoref{sec:RQ3}. These strategies underscore the adaptability of our system design, accommodating the diverse usage habits and preferences of individual users while achieving the goal of mitigating inconsistent decision outcomes.

\subsubsection{Effects on participants' cognitive workload}
\par
\begin{figure*}[htb]
\begin{centering}
\vspace{-1mm}
\includegraphics[width=\linewidth]{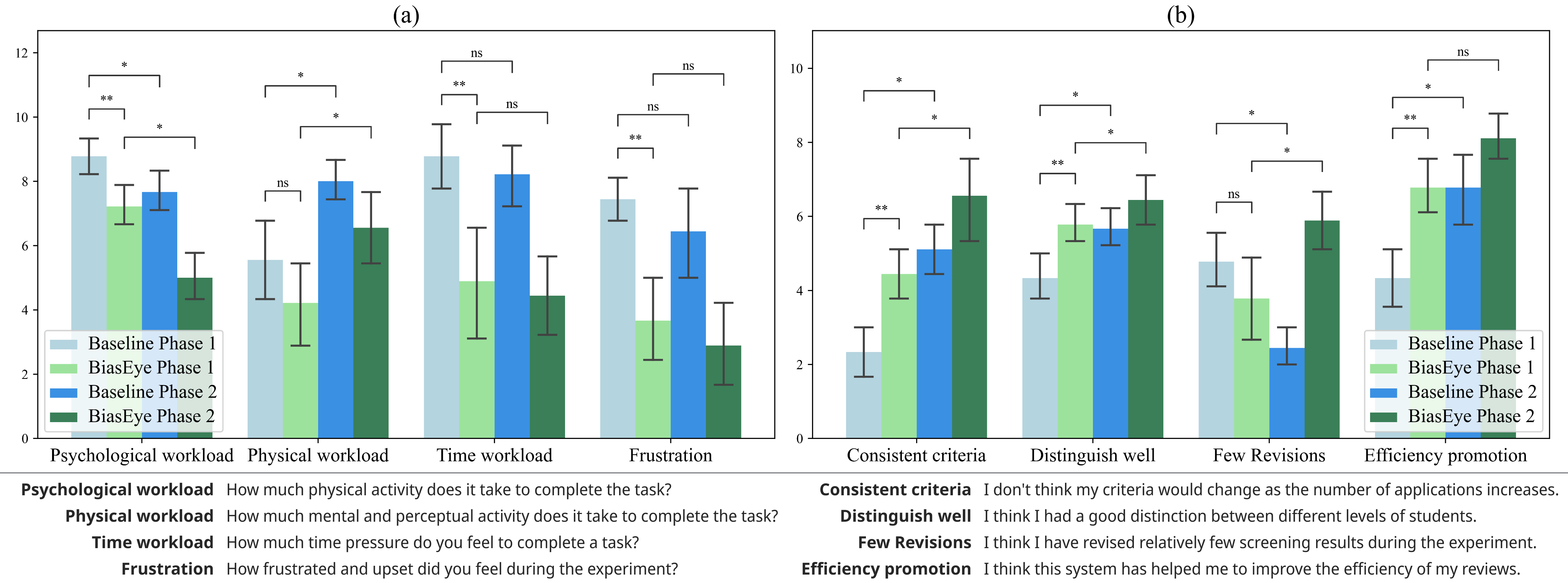}
\end{centering}
\vspace{-6mm}
\caption{Results of the (a) workload assessment and (b) self-evaluation in the questionnaire. Error bars indicate standard errors. (ns: p < .1; $^{\ast}$: p < .05; $^{\ast\ast}$: p < .01).}
\vspace{-3mm}
\label{fig:workloadEvaluation}
\end{figure*}

 In this section, we employ questionnaire data to assess the variations in workload among participants when comparing the Baseline and \textit{BiasEye} systems. The results are visually presented in \autoref{fig:workloadEvaluation}~(a). During Phase 1, \textit{BiasEye} significantly reduced psychological ($U=70.0, p < 0.01$) and time workload ($U=72.0, p < 0.01$) compared to the Baseline. Transitioning from Phase 1 to Phase 2, the introduction of \textit{Summary} page resulted in significant reductions in both psychological ($T=0.0, p < 0.05$ in Baseline, $T=0.0, p < 0.05$ in \textit{BiasEye}) and physical workloads ($T=0.0, p < 0.05$ in Baseline, $T=5.5, p < 0.05$ in \textit{BiasEye}) for both systems. Participants did not report significant changes about time workload and feeling of frustration.

\subsubsection{Effects on participants' self evaluation}
\par \autoref{fig:workloadEvaluation}(b), we present the differences in self-evaluation between the Baseline and the \textit{BiasEye} system. During Phase 1 of the experiment, participants using \textit{BiasEye} reported experiencing more consistent criteria ($U=8.0, p < 0.01$) and better distinction among applications ($U=11.5, p < 0.01$) compared to the Baseline group. Additionally, \textit{BiasEye} significantly improved the screening efficiency ($U=6.0, p < 0.01$). Moving on to Phase 2, the introduction of the \textit{Summary} page had a notable impact on both groups. It enhanced the criteria consistency ($T=0.0, p < 0.05$ in the Baseline group and $T=3.0, p < 0.05$ in the \textit{BiasEye} group) and improved the distinction among applications ($T=0.0, p < 0.05$ in the Baseline group and $T=0.0, p < 0.05$ in the \textit{BiasEye} group). However, it's important to note that only the Baseline group reported a significant increase in efficiency.

\subsection{RQ3. How will participants trust and collaborate with the ML method?} \label{sec:RQ3}

\par Through a qualitative analysis of video transcripts, we identified varying levels of trust among participants in the suggestions provided by the model. This trust, in turn, influenced their collaborative interactions with the machine learning-supported assistant system. Among the 19 participants in our study, only two opted not to utilize the model. The remaining 17 participants all made revisions based on the model's recommendations. It's important to note that participants retained ultimate decision-making authority when it came to screening results. They determined whether to accept, refer to, or question the prediction scores of an application, integrating their own understanding of the application materials. The machine learning method served as a supplementary tool, offering a clear and expedited path to identify inconsistencies within specific applications. More specifically, our study revealed the following findings into participants' trust in and collaboration with the ML method.

\par \textbf{Finding 9: Participants' trust in the model's performance is screening section-independent.} Participants' lack of trust in the model's performance in one section did not influence their trust in other sections. For instance, P6 remarked, ``\textit{The model's predictions in the \Com{Ho} section are not accurate, but it does help me identify many incorrect scores in the \EB{EB} section.}'' Similarly, P11 expressed, ``\textit{I'm quite confident in how the model handles quantitative data, but the content in the \ExA{EXA} section encompasses various elements, and I doubt the model could comprehend my criteria.}''

\par \textbf{Finding 10: Participants generally attempted to comprehend the rationale behind model predictions, but success was not guaranteed.} Participants often inferred the reasons behind the model's predictions by examining various factors, including the prediction itself, attributes with significant weights displayed on the \textit{Notification Card}, and raw information about each student. These inferences ranged from grasping the overall logical reasoning of the model to providing individual explanations for specific application predictions. For example, P3 remarked, ``\textit{Attributes on the notification involve scores of the English proficiency test (CET) and school ranking, but the model thinks I gave high scores for many applications... um, it is sensitive to CET scores, I care less unless the CET score is under $500$.}'' Conversely, according to P13, ``\textit{The model scores 2, but he comes from an experimental class at a university, with understandably low ranking that the model might have overlooked. I'm sticking to my opinion.}'' As the system did not explicitly specify the concrete attributes contributing to each application's prediction, there were instances where participants found it challenging to make successful inferences, leading to comments such as, ``\textit{I can't understand},'' as noted by several participants.

\par \textbf{Finding 11: Participants tended to question the rationality of their decisions when there was a significant disparity between the predictions and human scores/expectations.} Participants' awareness of these differences stemmed from two primary sources. On one hand, it was influenced by the overall color trend of the ID text in the \textit{Comparison} view, as noted by P16 who mentioned, ``\textit{There are many red colors, and I am overwhelmed.}'' On the other hand, participants observed discrepancies between the human scores they assigned and the predicted scores for each application. P8 remarked, ``\textit{The prediction is around 5 points, but why did I only give 1 point? Although I don't quite understand why it scores 5, I decide to increase the score a bit.}'' Despite being informed at the beginning of the experiment that the model's predictions may be inaccurate, participants still exhibited a degree of blind belief and reliance on the model, particularly when they felt uncertain about an application. As P8 questioned, ``\textit{I am struggling with this score, or should I listen to the model?}'' Nevertheless, it's worth noting that the proposed system has mitigated confirmation bias to some extent by encouraging participants to engage in a second round of deliberation.

\par Through an analysis of the video transcripts, we also identified various factors that influenced participants' trust.
        
\par \textbf{Factor 1: The consistency between keywords and participants' perception of decision criteria.} The attributes listed on the \textit{Notification Card} served as the initial point for participants to grasp the model's functioning. When these attributes did not align with the participants' preconceived criteria, it led to doubts regarding the model's predictions. For instance, P11 exhibited skepticism towards the attributes in the \ExA{ExA} section. Upon identifying discrepancies and disagreeing with the predictions for three applications in the \textit{Comparison} view, P11 promptly cross-referenced and verified the scores of multiple applications in the \textit{Ex-situ Table} independently. We observed that inconsistent perceptions could also arise from differences in how attributes were categorized and participants' mental frameworks. For instance, competitions were initially categorized into different subjects within the \Com{Com} section. However, participants were often unaware of the distinctions between different subject areas within competitions, particularly for competitions that were rarely mentioned in the materials and thus not well-remembered or paid attention to.

\par \textbf{Factor 2: The significance of differences between predictions and human scores.}Participants exhibited strong trust in prediction scores that closely matched or were consistent with human scores. None of the participants actively sought applications with \closeto{gray-colored} IDs in the \textit{Comparison} view or those where human scores and predictions were sorted in the same order in the \textit{Ex-situ Table}. As P9 stated, ``\textit{Both the model and I agree with these scores, so there's no issue at all.}'' The greater the difference between the human score and prediction, the more likely it was for inconsistencies in applications to exceed the participants' threshold and capture their attention. For instance, P6 commented, ``\textit{Differences less than one are not an issue. I'll check the others that had larger score differences.}'' Conversely, the overall color trend of IDs in the \textit{Comparison} view also influenced participants' trust in the model's predictions. P12 mentioned, ``\textit{There are many \closeto{gray} ones, so I believe that the model has learned well.}'' It's essential to note that this factor does not contradict the phenomenon of self-doubt arising from \higherthan{higher}/\lowerthan{lower} color trends mentioned in Finding 11.

\par \textbf{Factor 3: The presence of sufficient evidence for confirmation and trust.} Participants were more inclined to trust predictions when they discovered ample evidence to support them. This evidence could be gathered by verifying whether key information in an application had been overlooked or by making comparisons between multiple applications. For example, P4 admitted, ``\textit{It's my fault. I didn't pay attention to the Mathematical Contest In Modeling just now.}'' In a similar vein, P11 commented, ``\textit{Compared to other applications that meet my expectations of four, this application is indeed slightly worse. I will follow the model and adjust it to three.}'' Differences in how participants and the model interpreted the same information sometimes hindered their trust in the predictions. For example, P13 from Group A, which did not have access to the \textit{Statistical} view, questioned, ``\textit{Why did the model give him a score of four when he's from an average university and his GPA is not at the top level?}'' Subsequently, the participant referred to supplementary materials and found that the university was ranked around the top 50, which is considered quite good. This incident highlights how human judgment can be influenced by personal experiences, potentially leading to biases, such as the availability heuristic~\cite{Tversky:1974:Judgment} and confirmation bias, which makes individuals ignore objective truths. Notably, this issue was not observed among participants in Group B (\textit{BiasEye}), as the \textit{Statistical} view provided valuable evidence.

\par \textbf{Factor 4: Participants' intrinsic perceptions of machine learning.} Participants' intrinsic beliefs about machine learning significantly influenced their trust in the system. For instance, P7 expressed confidence, stating, ``\textit{The system is definitely more accurate than I am.}'' Similarly, P15 held the view that, ``\textit{Machines don't get tired; they have no blind spots in attention.}'' Conversely, some participants like P4 were more skeptical, stating, ``\textit{I've learned about machine learning algorithms. If some attributes do not appear in the selected samples, it cannot learn them.} P18 also struck a balance, noting, ``\textit{I believe that machine learning can assist me, but I'm aware it has limitations too. I won't blindly follow it.}'' These inherent perceptions of machine learning played a pivotal role in shaping participants' trust.

\vspace{-2mm}
\section{Discussion and Limitation}
\par In this section, we extract future design considerations DC1$\sim$4 (\autoref{sec:discussDC}) from our analysis results and questionnaire feedback. We also explore potential generalizations of our findings to other domains in \autoref{sec:discussGen} and reflect on the limitations of our work in \autoref{sec:discussLimit}.

\subsection{Design Consideration} \label{sec:discussDC}
\par \textbf{DC1: Improve the interactive capability of the system.}
Participants appreciated \textit{BiasEye}'s interactive features in our study, such as real-time score box-plot updates, highlighting the current application in the \textit{Statistical} view, and quick navigation between \textit{Screening Sheets}, which alleviate their workload. A bias-aware intelligent interface for decision-making should seamlessly incorporate interactive functionality, enabling users to devote more cognitive resources to thoughtful judgment. This integration is essential for encouraging users to actively address biases in decision outcomes. Additionally, such systems should gather and present more contextual information to support well-informed decisions. Our study revealed that certain participants in group A, like P5 and P13, infrequently referred to supplementary materials and were influenced by personal experiences, leading to inconsistent screening results. To alleviate the impact of inadequate or incorrect memory and perception, a recommendation is to implement dynamic annotations within the interface. These annotations could include hyperlinks to pertinent information such as school, major, competition details, and data on past admitted students. If this information could be aggregated, the interface might visualize a comparison between individual and collective data. Consequently, instead of facing unfamiliar and ambiguous perceptions, users could swiftly grasp relevant information.

\par \textbf{DC2: Simplify views and visual designs.} 
The design of visualization and functionality should prioritize intuitiveness, avoiding the need for complex computer expertise and minimizing the learning curve. In our study, participants acknowledged the attractiveness of glyphs but found their placement lacked meaning, as highlighted by P6 and P11. The process of visualized dimensionality reduction added cognitive demands and had the potential to cause misunderstanding. Interestingly, the \textit{Ex-situ Table} view was deemed more user-friendly than the \textit{Comparison} view, leading participants to prefer a format combining glyphs with a table presentation. As a result, future interface designs could incorporate tables with multiple straightforward mini-charts, offering a more effective way for users to understand data without increasing cognitive load. Additionally, for complex decision tasks like material screening, it remains uncertain whether a multi-view visual analysis strategy is a more effective option.
    
\par \textbf{DC3: Enhance machine learning with human guidance.}
Our observations unveiled that pre-specified model training attributes approximated only a limited subset of participants' personal criteria. Despite some commonalities, each participant had unique focus areas. A universal model struggled to differentiate individual applications based on specific criteria and often misclassified similar applications due to attribute redundancy. While more intricate models and comprehensive attributes could align better with actual screening criteria, there exists a trade-off between a perfect fit and real-time response. AI methods may not be as proficient or accurate as domain experts in verifying applicants' contributions and identifying potential exaggerations. Moreover, AI faces challenges in acquiring contextual knowledge, such as how personal experiences are influenced by socioeconomic and geographic disparities. Implicit discrimination may be hidden in the superficial quantification of applicants based on factors like SAT scores and academic awards. To address the limitations of ML methods, future intelligent screening systems should adopt ``human-in-the-loop'' approaches. Specifically, the interface can allow model training for customized attributes, correction of deviant model, special marking and score lock of outliers (e.g., students at risk of fraud or those considered deserving of preferential treatment).

\par \textbf{DC4: Acknowledge the constraints of AI assistance techniques.}
The majority of participants (11 out of 18, with 5 not providing a response) acknowledged that automated information extraction improved retrieval efficiency and reduced their workload. Additionally, they found that the ML method assisted in addressing inconsistencies in screening decisions. However, it was also observed that participants tended to heavily rely on AI support methods, particularly the ML predictions. Consider the limitations of ML methods mentioned in DC3, human-machine collaboration strategies should be devised to promote AI in complement with human decision-making, rather than allowing unchecked dependence on algorithms. In this context, future intelligent interfaces should discourage the outright use of AI in initial decision-making, instead supervising users to adopt recommendations with adequately understanding. For example, system can pop up temporary windows to declare the limitation of the AI method, inquire about users' confidence in their personal judgment versus AI prediction, and encourage users to assess the consistency of their judgment with AI recommendations.

\vspace{-2mm}
\subsection{Generalizability} \label{sec:discussGen}
\par Tasks such as corporate hiring, fund applications, and scholarship selections often require the evaluation of numerous multi-dimensional and multi-modal materials. These tasks commonly face different cognitive biases, resulting in inconsistent outcomes and affecting individual fairness. \textit{BiasEye} is flexible and can be tailored by modifying the necessary attributes and algorithms to meet the specific requirements of a task. In our user studies, the simple \textit{Ranking SVM} demonstrated encouraging results in assisting with bias mitigation. We are also interested in exploring more advanced approaches, such as neural networks, capable of capturing complex reasoning processes to further improve the effectiveness of bias mitigation.

\vspace{-2mm}
\subsection{Limitation} \label{sec:discussLimit}
\par This study primarily evaluates our bias-aware screening system design, excluding information extraction as an experimental condition. However, it's important to note that data extraction and classification models can introduce errors, highlighting the need for better document organization in application submissions. We recommend institutions implement formal systems for collecting structured personal information alongside documents, which can improve screening system design and functionality. Additionally, \textit{BiasEye} relied on quantitative attributes for prediction, potentially missing nuanced human screening preferences, especially for indicators like project content and quality. To address this, exploring specialized language models or textual information extraction features may enhance learning and prediction, particularly in detecting biases in PSs and LoRs. Future systems could also simplify screening through content analysis for categorical comparisons of applicants. Lastly, due to constraints, we conducted a controlled in-lab study with senior students, not directly comparable to expert admissions reviewers. We plan to pursue a field study after further system optimizations.

\vspace{-2mm}
\section{Conclusion and Future Work}
\par This study introduces \textit{BiasEye}, a specialized interactive system designed to address, detect, and mitigate potential biases in real-time screening processes. \textit{BiasEye} provides users with clear global views of information, aiding in fair screening criteria formulation. It also helps identify biases by comparing actual rankings with model-predicted ones, offering immediate means for adjustment. Results from a user study show that \textit{BiasEye} significantly improves reviewers' decision-making by visualizing potential biases, suggesting its value across screening tasks. Future improvements may involve advanced machine learning algorithms and broader domain applications, including enterprise and government contexts. \textit{BiasEye} development could inspire more tools for impartial decision-making and bias reduction.

\vspace{-2mm}
\begin{acks}
We would like to express our gratitude to our domain experts and the anonymous reviewers for their insightful comments. This work is funded by grants from the National Natural Science Foundation of China (No. 62372298), the Shanghai Frontiers Science Center of Human-centered Artificial Intelligence (ShangHAI), and the Key Laboratory of Intelligent Perception and Human-Machine Collaboration (ShanghaiTech University), Ministry of Education.

\end{acks}

\vspace{-2mm}
\bibliographystyle{ACM-Reference-Format}
\bibliography{ref}

\appendix

\begin{flushleft}
    \Large
    \textbf{\textsf{Appendix}}
\end{flushleft}

\vspace{-2mm}
\section{Baseline Design}

\begin{figure}[ht]
\centering
\vspace{-3mm}
\includegraphics[width=0.82\linewidth]{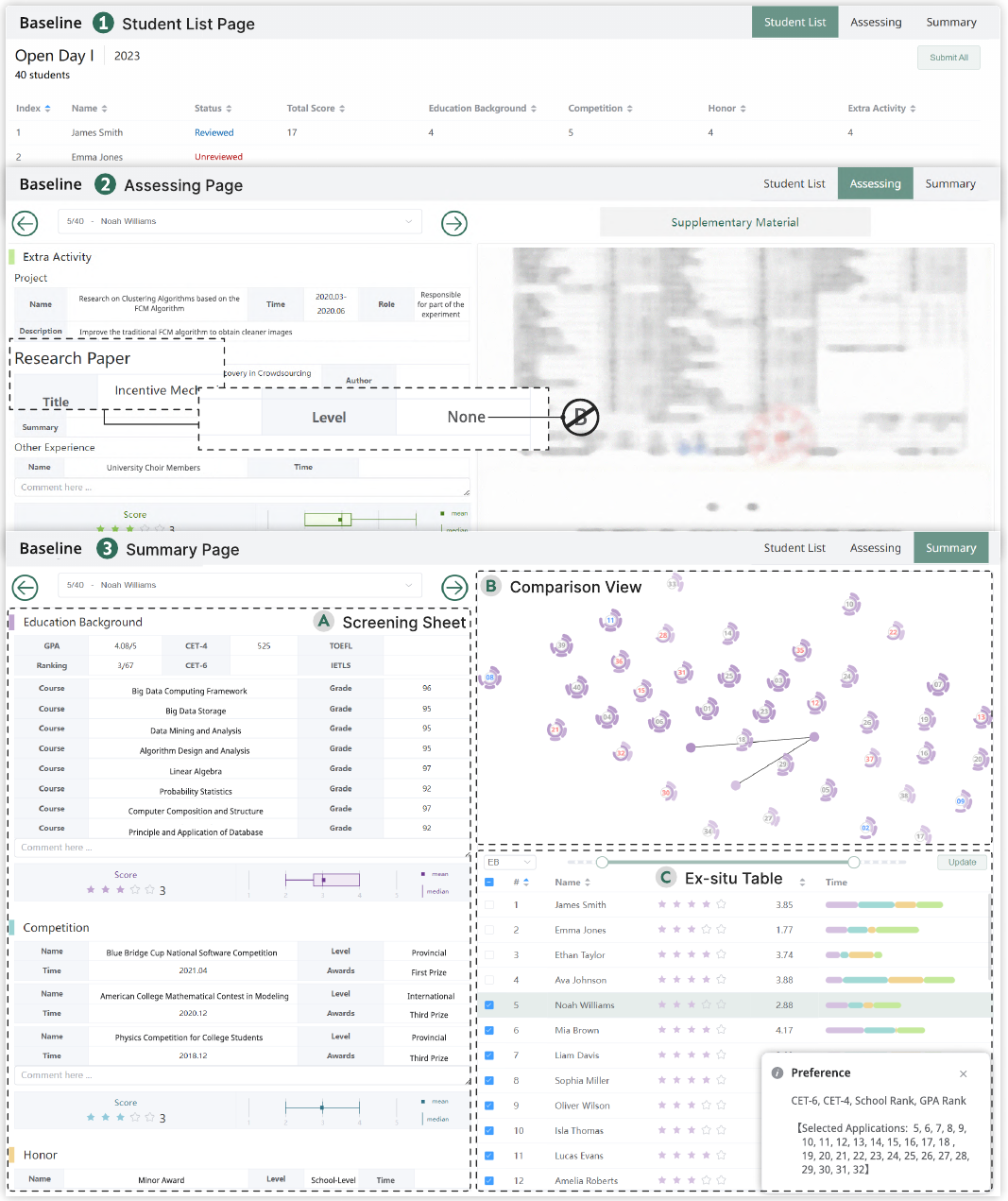}
\vspace{-2mm}
\caption{The baseline system used in the experiment. In Phase I, participants are limited to using only (1) the \textit{Student List} page and (2) the \textit{Assessing} page. In Phase II, participants can also utilize (3) the \textit{Summary} page.}
\label{appendix:baselineSystem}
\vspace{-3mm}
\end{figure}

\section{Usage Scenario}
\par Let's consider Professor Alex, 
tasked with reviewing a large number of university admissions applications each May, faced the challenge of juggling this with his teaching and research duties.
He found that meticulously reading through a large volume of application information consumed a significant amount of his energy. To cope with this, he resorted to manually extracting essential information from the PDF files, enabling him to make comparisons between applications before and after this data retrieval step, ultimately aiding in his decision-making process. While these heuristics proved effective in most cases, they did carry a risk of errors, especially when Alex began to experience fatigue. However, considering the significance of college admissions as a crucial societal matter, Alex dedicated a substantial amount of time to meticulously review his screening results before submission. He did this out of concern that he might overlook outstanding applicants in the process. This particular aspect of the procedure heavily relied on Alex's subjective judgment.

\par This year, Alex utilized \textit{BiasEye} for the application screening process. He began by reviewing the \textit{Statistical} view to conduct an initial evaluation of the applicants before commencing the screening. As the process continued, he increasingly relied on this view to recall and grasp quantitative information, such as school rankings. \textit{BiasEye} conveniently extracted information according to sections in advance, effectively conserving Alex's energy. After evaluating approximately forty to fifty applications, Alex proceeded to the \textit{Summary} page to examine his screening outcomes. Initially, he noticed some conspicuous irregularities in time allocation and identified an incorrect score in the \ExA{Extra Activity} section of one application. He realized that he had misread some vital information during his hurried decision-making process. Subsequently, Alex adopted a systematic approach to reviewing each section. He began by selecting the education section from the drop-down menu. Recognizing that he might have been too cautious initially due to a lack of confidence and that his decision-making quality had decreased towards the end of the process due to fatigue, he adjusted the slider to focus on the middle subset of applications that aligned with his screening preferences. He carefully inspected the contents on the Notification Card and found that the attributes aligned with his expectations. Alex noted that the \textit{Comparison} view effectively highlighted discrepancies between the model and human `decisions', making it easier to identify applications with lower or higher scores based on the color of the IDs. However, he felt that he should prioritize applications with a significant disparity between the inner and outer arcs. Additionally, by examining the center dots, he realized that he had only assigned ratings ranging from 2 to 5. While reviewing application $\#11$, Alex observed that the score was lower than the prediction. In the \textit{Comparison} view, he discovered that this application closely resembled $\#14$, indicating a degree of similarity. Alex conducted a detailed comparison between these two applications and decided to follow the model's suggestion by modifying the score to 3. Employing a similar approach, he made adjustments to more scores. In the end, Alex's confidence in his screening results grew, and he was pleased to have mitigated the inconsistencies in his screening decisions.

\end{document}